\documentclass[10pt,journal,final]{IEEEtran}
\usepackage{amsmath}
\usepackage{xcolor}
\usepackage{mathtools}
\DeclarePairedDelimiterX{\norm}[1]{\lVert}{\rVert}{#1}
\usepackage[english]{babel}
\usepackage[utf8]{inputenc}
\usepackage{algorithm}
\usepackage{algpseudocode}
\usepackage{stackengine}
\setstackEOL{\\}
\makeatletter
\usepackage{cite}
\usepackage{amsmath,amssymb,amsfonts}
\usepackage{graphicx}
\usepackage{textcomp}
\usepackage{tabularx}
\usepackage{makecell}
\usepackage{diagbox}
\usepackage[long]{optidef}
\usepackage{multicol, blindtext}
\usepackage{soul}
\usepackage{lipsum,adjustbox}
\usepackage{pgfplots,pstricks,tikz}
\usetikzlibrary{calc}
\usetikzlibrary{shapes,decorations,arrows,backgrounds}
\usepackage{pgfplots}
\usepgfplotslibrary{groupplots}
\usepackage{xcolor}
\usepackage{multirow}
\definecolor{mytextcolor}{rgb}{1,0,0}
\setstcolor{red}
\definecolor{mac_color}{RGB}{90,11,47}
\definecolor{myblue}{RGB}{168,255,255}
\definecolor{mygray}{RGB}{192,192,192}
\definecolor{mygray2}{RGB}{70,70,70}
\definecolor{myred}{RGB}{255,102,102}
\colorlet{mac_light}{mac_color!5}
\colorlet{mac_med}{mac_color!55}
\colorlet{mac_backshade}{mac_color!1}

\newtheorem{proposition}{Proposition}

\newtheorem{remark}{Remark}

\newtheorem{definition}{Definition}
\begin{document}
\title{Distributed Cooperation Under Uncertainty in Drone-Based Wireless Networks: A Bayesian Coalitional Game}
\author{Vandana Mittal, Setareh Maghsudi, and Ekram Hossain, \IEEEmembership{Fellow, IEEE}
\thanks{Vandana Mittal and Ekram Hossain are with Department of Electrical and Computer Engineering, University of Manitoba (E-mail: mittalv@myumanitoba.ca, ekram.hossain@umanitoba.ca). Setareh Maghsudi is with the Department of Electrical Engineering and Computer Science, Technical University of Berlin, 10623 Berlin, Germany (E-mail: maghsudi@tu-berlin.de).}}
\maketitle
\begin{abstract}
We study the resource sharing problem in a drone-based wireless network.
We consider a distributed control setting under uncertainty (i.e. unavailability of full information).  In particular, the drones cooperate in serving the users while pooling their spectrum and energy resources in the absence of prior knowledge about different system characteristics such as the amount of available power at the other drones. We cast the aforementioned problem as a Bayesian cooperative game in which the agents (drones) engage in a coalition formation process, where the goal is to maximize the overall transmission rate of the network. The drones update their beliefs using a novel technique that combines the maximum likelihood estimation with Kullback-Leibler divergence. We propose a decision-making strategy for repeated coalition formation that converges to a stable coalition structure. We analyze the performance of the proposed approach by both theoretical analysis and simulations.
\end{abstract}
\section{Introduction}
\label{sec:introduction}
The current wireless networking paradigm makes a significant step towards building up the ultra-reliable, low latency, power and spectral efficient communication with the help of technologies such as millimeter wave and massive MIMO communications, seamless integration of licensed and unlicensed bands, as well as intelligent spectrum usage and management \cite{giordani2020toward}. Nonetheless, the rapid increase of wireless traffic calls for further enhancement through the integration of aerial base stations (e.g. drones) in order to improve the coverage, capacity, and connectivity of existing terrestrial cellular networks. The potential of using drones as aerial base stations stems from their altitude flexibility and the possibility of establishing the line-of-sight (LoS) link towards ground users. Despite the huge potential, drone-based wireless communication systems face a variety of challenges including precise channel modeling, efficient users-to-drones association, interference management, trajectory optimization, and resource management and control, etc. \cite{sekander2018multi}. 

In general, centralized management and control of a network requires the availability of the global network information at a central controller that performs network optimization. These methods, however, suffer from excessive overhead and computational cost. From this point of view, it is important to develop distributed solution methods that are robust to uncertainty and information shortage. One mathematical tool that enables rigorous analysis of multi-agent systems under uncertainty is the cooperative game theory. An example is \cite{maghsudi2017distributed} that investigates distributed task management in networked cyber-physical systems.

Based on the flying mechanism of drones, they are broadly classified into two categories as fixed-wing drones and rotary-wing drones \cite{sekander2020statistical}, \cite{zeng2016wireless}. Fixed-wing drones usually can carry heavy payload and can travel at a high speed. However, they need to maintain continuous forward motion to remain aloft; Therefore, they are not suitable for stationary applications. Rotary-wing drones, despite having limited payload capacity and mobility, can either move in any direction or stay stationary in the air. Thus, such drones can be used as hovering drones at a certain location to ensure continuous coverage. However, they consume a significant amount of power to keep them hovering in the air all the time. A particular type of drone is selected depending on the application. In this paper, we consider the rotary-wing hovering drones, each placed at a fixed location to provide maximum coverage to its assigned users. We assume that the available power at each drone is time-varying. This variation in the available power in a  drone is due to the fact that the power supply and energy dissipation at each drone can vary depending on its type, mode of operation, location during communication (and hence propagation condition), etc. As a specific example, a drone with energy harvesting capability  can harvest energy from stochastic resources and then communicate with the ground users or base stations by using the harvested energy~\cite{sekander2020statistical}. Since energy harvesting can be intermittent and uncertain, the amount of available energy will be statistically varying.


In the setting described above, we address the distributed resource sharing problem, where there is some uncertainty about  network parameters such as the power availability in each drone. More specifically, to enhance network performance, the drones can make cooperative clusters. The drones within a cluster can share their available power and spectrum resource to optimize the service provided to the users. Given no prior information about the statistical characteristics of the available power, the goal is to find an optimal structure of drone clusters along with the best channel assignment and power allocation to each user in the cluster such that the overall transmission rate (social welfare) in the network is maximized.

To solve the described problem, we formulate it as a Bayesian Coalition Formation Game (BCFG), where the drones represent the players. Each drone, being uncertain about the available power, or the \textit{type}, of others, forms some belief. Given the beliefs, the drones engage in a coalition formation process that aims at optimizing the service for the users assigned to the coalition members. By observing the outcome of its action and possibly some side-information, each drone then updates its belief. The process continues until convergence. The main contributions of the paper include the following:
\begin{itemize}
\item For a drone-based wireless network, we formulate the resource sharing problem under uncertainty as an optimization problem. The joint optimization problem includes (i) learning he \textit{types} along with finding the optimal cluster configurations of the drones, (ii) user-channel assignment, and (iii) power allocation in each cluster. Since this is a combinatorial problem, we divide the optimization problem into three sub-problems. We cast the  cooperation problem among drones as a Bayesian coalition formation game. We propose a distributed approach based on best-reply dynamics to obtain the optimal clusters of drones. In each cluster, the user-association and channel allocation problem is formulated as a bipartite matching problem. Finally, we use a water-filling algorithm for power allocation.
\item For numerical analysis, we simulate a variety of network settings. We investigate the performance of our proposed solution, as well as the effect of uncertainty and cooperation, by implementing several benchmarking approaches. These include the baseline configuration in which the drones perform independently without cooperation, distributed coalition formation with full information, and the social optimal case. The results establish the superior performance of the proposed scheme in terms of improvement in the sum rate of the network and individual rate of the drones. We also show the effect of overlap in the distribution of the {\em types} on convergence.
\item Concerning the methodology of Bayesian coalition formation, we propose a novel belief-updating method. More precisely, instead of updating the belief using Bayes' rule, our approach first estimates  the parameters of the distribution (of {\em type}) and then finds the closeness of the estimated parameters with the given set of types followed by averaging. We also use local information to update the belief; that is, the drones share the information only inside the corresponding coalition, thereby reducing the feedback and signaling overhead.
\end{itemize}
Compared to the state-of-the-art, our approach offers the following advantages:
\begin{itemize}
\item It is more scalable since we assume limited use of a central controller (e.g. a software-defined controller) in the cooperation process or the availability of precise information about the critical variables. 
\item It can be used for networks where the users or drones leave/join the network dynamically. Upon detection of the changes in the network dynamics or type sets (e.g. by the controller), the algorithm can be triggered to execute using the updated information.
\item The proposed coalition formation model and decision-making strategy do not depend on the subsequent resource allocation or the statistical characteristics (such as the distribution) of the available power at the drones. Therefore, it is highly adaptable to different systems beyond the drone networks. 
\end{itemize}

The organization of the paper is as follows. \textbf{Section \ref{Sec:SOA}} provides a brief overview on the state-of-the-art research. \textbf{Section \ref{system_model}} describes the system model. We formulate the resource sharing problem in \textbf{Section \ref{subsec:Optimization}}.  In  \textbf{Section \ref{sebsec:BCFG}}, we model the formulated problem as a  Bayesian coalition formation games, and \textbf{Section \ref{solution_approach}} includes the algorithmic solutions. \textbf{Section \ref{anaysis}} presents the theoretical analysis. In \textbf{Section \ref{results}}, we evaluate the performance numerically. \textbf{Section \ref{conclusion}} concludes the paper.
\section{Related Work}
\label{Sec:SOA}
Resource management and energy efficiency for drone-assisted cellular wireless networks have attracted significant attention of the research community. Reference \cite{tran2017cooperative} proposes an energy-efficient scheduling framework for cooperative drone communication. In \cite{ceran2017optimal}, the authors propose an optimal resource allocation strategy for an 
energy-harvesting flying access point. Reference \cite{mozaffari2017wireless} studies the problem of flight time optimization and bandwidth allocation of the drones that serve ground users. The authors of \cite{chen2017liquid} propose a resource allocation framework for cache-enabled drones that provide services to the ground users over unlicensed and licensed bands. Reference \cite{sharma2016uav} studies optimal user and drone assignment for capacity improvement in drone-assisted heterogeneous wireless networks. In \cite{kalantari2016number}, the authors jointly optimize the number of active drones and their locations to maximize the coverage. In \cite{cui2019multi}, the authors propose a framework based on stochastic game theory to optimize the performance of multi-drone networks by a joint selection of power levels, sub-channels, and users. The authors in \cite{lagum2017strategic} propose a novel method for the strategic placement of multiple drones along with base stations in a large scale network.  Reference \cite{kalantari2017backhaul} studies the backhaul-aware optimal placement of drones and base stations to maximize the number of served users. Reference \cite{ghazzai2018trajectory} develops a framework consisting of swarms of UAVs as flying relays for delay-intolerant and bandwidth-hungry applications. Reference \cite{samir2019trajectory} presents a method for resource allocation and trajectory planning for multiple UAVs that deliver data in vehicular networks. Reference \cite{8674587} studies the positioning of nodes in a dynamic UAV swarm network with the goal of optimal throughput communication. In \cite{ye2019joint}, the authors develop a framework for UAV-enabled wireless-powered Internet of Things. They maximize the sum throughput of the network by joint optimization of UAV placement, time allocation, and the UAV-device association. Reference \cite{dinh2019joint} investigates joint optimization of UAV location planning, content placement, transmit beamforming, and user admission decisions. The objective is to maximize the number of served users while satisfying the minimum rate requirements, where the UAVs have limited storage capacity. Authors of \cite{zhang2019framework} study drone-mounted in-band full-duplex heterogeneous networks. They maximize the network's transmission performance by joint optimization of drone placement, power and bandwidth allocation, and user assignment.

Beyond drone networks, there is a large body of literature that study the resource allocation and user association problems in heterogeneous networks. For instance, \cite{fooladivanda2012joint} formulates a centralized framework to analyze and compare various user association-, resource allocation-, and interference management schemes. Similarly, the authors in \cite{wang2017joint} study joint optimization of user association, power allocation, and channel allocation in multi-cell multi-association OFDMA heterogeneous networks. They decompose the problem into two subproblems. The sub-problems are then solved alternatively to obtain the local optimal solution. Reference \cite{zhao2019deep} develops a multi-agent reinforcement learning-based distributed solution.

From the methodology perspective, coalition formation game without uncertainty has been a popular tool to solve wireless communication problems. For example, in \cite{saad2009distributed}, the authors propose distributed cooperation among single antennas to form the virtual multi-antenna system to improve network performance. In \cite{gharehshiran2012collaborative}, users collaborate for sharing the sub-channels in the cognitive LTE femtocells environment. Cooperation among femtocell access points by sharing the excess computational resources is proposed in \cite{ tanzil2016distributed}. The goal is cost- and delay reduction by avoiding unnecessary offloading to the remote cloud. References \cite{wu2011coalition}, \cite{wu2012cooperative}, and \cite{gharehshiran2010coalition} study the application of coalition formation game in wireless sensors networks. Specifically, \cite{wu2011coalition} considers the network's lifetime expansion with the desired quality of service requirements. Reference \cite{wu2012cooperative} proposes to balance the energy efficiency and QoS provisioning for cooperation in a clustered wireless sensor network. Reference \cite{ren2017coalition} shows the application coalition formation game for cooperative networks with simultaneous wireless information and power transfer.           
A few papers use coalition formation game with uncertainty to solve the resource allocation problem in wireless networks. For example, the authors of \cite{akkarajitsakul2011coalition} use the Bayesian coalition game with nontransferable utility (NTU) for packet delivery among mobile nodes under uncertainty in node behavior. Reference \cite{asheralieva2017bayesian} proposes utilizing coalition formation based on Bayesian reinforcement learning for distributed resource sharing in device-to-device (D2D) enabled heterogeneous cellular networks. Similarly, \cite{xiao2015bayesian} presents a Bayesian overlapping coalition formation game for spectrum sharing between multiple co-located cellular networks and a set of D2D links. Reference \cite{khan2010modeling} models the dynamics of coalition formation games for spectrum sharing in an interference channel. Reference \cite{Maghsudi19:DTM} studies the cooperative task allocation problem among heterogeneous cyber-physical systems under uncertainty about the randomly arrived tasks and stochastic systems' types. The authors model the problem as a multi-state stochastic cooperative game with state uncertainty. Reference \cite{Maghsudi17:DUA} solves the cooperative user association problem under uncertainty using an approach based on the exchange economy. 

None of the aforementioned works investigate the problem of distributed cooperation among drones for  resource sharing under uncertainty considering realistic network and channel propagation conditions. 
\begin{table*}[!t]
\centering
\caption{Important notations and definitions}
\begin{tabular}{|l|l|}
\hline
\textbf{Notation}  &  \textbf{Definition}  \\
\hline
$\mathcal{D}$, $\mathcal{N}$, and $\mathcal{Q}$ & Set of drones, users, and channels \\
$\mathcal{D}^{C_k}$, $\mathcal{N}^{C_k}$, and $\mathcal{Q}^{C_k}$& Set of drones, users, and channels in coalition $C_k$ \\
$\mathcal{W}$ & Set of coalition structures for $D$ drones with cardinality $W$ \\
$w = \{ C_1, \dots ,C_k, \dots ,C_l \} $ & Coalition structure $w \in \mathcal{W}$ consisting of $l$ coalitions \\
$ w^* $ & Nash stable coalition structure \\
$\textbf{X}_w, \textbf{Y}_w$ & Set of channel and user- assignment matrices for coalition structure $w$ \\
$\textbf{p}^{C_k} = [p_1, \dots, p_n, \dots, p_{\mathcal{N}^{C_k}} ]$ & Power vector consisting of power of each user in coalition $C_k$ \\
$\bar{L}_{dn,q}$ & Average path loss of drone $d$ towards user $n$ over channel $q$ \\
$R_{dn}(q,p_n)$ & Transmission rate of user $n$ from drone $d$ over channel $q$ with power $p_n$ \\
$x_{qn}, y_{dn} $  & Binary variables with value $1$ if an assignment exists and $0$ otherwise \\
$P_d$  & Power of drone $d$ \\
$P^{C_k} = \sum_{d = 1}^{\mathcal{D}} P_d$ & Total power of the coalition $C_k$ \\
$\mathcal{T} = \{ T_1, \dots, T_M\}$ & Type set consisting of $M$ types \\
$t^d \in \mathcal{T}, P_d = \mathbb{E}(t^d)$ & Type of drone $d$, power of drone (expected value of its type) \\
$\textbf{T} \in \otimes_{d = 1}^{D}\{ T_1, T_2, \dots,T_M \}$ & Type space of $D$ drones \\
$\textbf{T}^{-d} \in \otimes_{x \in \mathcal{D} \backslash d}\{ T_1, T_2, \dots,T_M \}$ & Type space of $D \backslash d$ drones \\
$\textbf{t}^{-d} = [t^d_1, \dots,t^d_j \dots, t^d_{\mathcal{D}\backslash d}] \in \textbf{T}^{-d} $ & Belief space of drone $d$ \\
$\textbf{t}_{C_k} \in \textbf{T}_{C_k} = \otimes_{d \in C_k} \mathcal{T}$& Type vector of drones in coalition $C_k$ belonging to the type space $\textbf{T}_{C_k}$  \\
$\textbf{t}_{C_k}^{-d} \in \textbf{T}^{-d}_{C_k} = \otimes_{j \in \{ C_k \backslash d \} } \mathcal{T}$ & Type vector based on drone $d$ about the members $j \in \{ C_k \backslash d \}$ from type space $\textbf{T}^{-d}_{C_k}$ \\
$B(\textbf{t}_{C_k}^{-d})$ & Belief of drone $d$ about the type vector $\textbf{t}_{C_k}^{-d}$ \\
$P(\textbf{t}_{C_k}^{-d})$ & Power of coalition $C_k$ as a function of type vector $\textbf{t}_{C_k}^{-d}$ \\
$\mathcal{A}^{C_k}$ & Set of coalition actions consisting of all possible channel associations and power allocations \\
$\bar{q}_d^d, \bar{q}_j^d$ & Expected pay off of drone $d$ and drone $j$ based on drone $d$'s belief about members of $C_k$ \\
$\bar{q}_{total}$  & Total expected payoff of the network \\
$\textbf{W}$, $\rho_{w,w'}$  & Transition matrix of the Markov chain of BCFG, Transition probability from $w$ to $w'$ \\
$\pi_{w}$  & Formation probability for coalition structure $w$ \\
\hline
\end{tabular}
\label{table:notations and definitions}
\end{table*}
\section{System Model and Assumptions}
\label{system_model}
In this section, we describe the system model. \textbf{Table \ref{table:notations and definitions}} summarizes the most important variables that are frequently used in this paper.  
\begin{figure}[!htb] 
\label{network_figure}
\centering
\includegraphics[width=0.7\columnwidth]{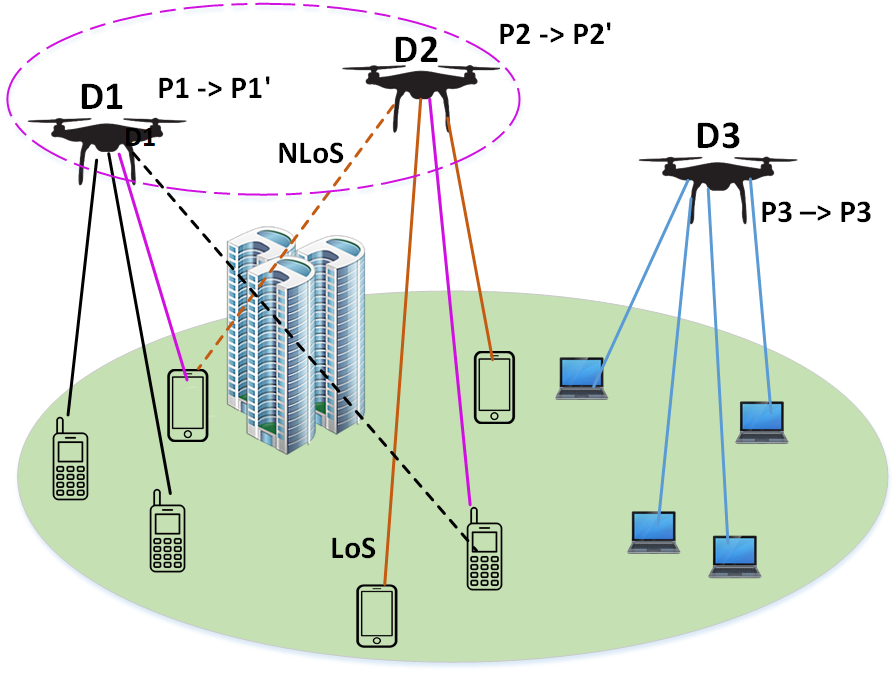}
\caption{Network model: Drones $D1$ and $D2$ cooperate by forming a coalition to ensure the possibility of line-of-sight (LoS) transmission, thereby improving the transmission performance. }
\label{Fig:system_model}	
\end{figure}
\subsection{Network Model}
\label{subsec:NetModel}
As shown in \textbf{Fig. \ref{Fig:system_model}}, we consider a drone network with multiple hovering drones with fixed locations. Each drone shall serve a pre-defined set of users that are distributed uniformly over a wide geographical area. This happens, among others, when the drones belong to different service providers, or when the drones support the overload traffic from the terrestrial cellular system. We represent the set of drones, channels, and users respectively by $\mathcal{D}=\{1,\dots,D\}$, $\mathcal{Q}=\{1,\dots,Q\}$, and $\mathcal{N}=\{1,\dots,N\}$. Each drone $d \in \mathcal{D}$ has access to $Q_d$ orthogonal channels, each with bandwidth $B$. The drone allocates each channel only to one of its assigned user. Therefore, the number of users that can be served by a drone at each time instant is at most equal to the number of available channels. Primarily, each drone operates independently of the others, routing the traffic of its own users only. We refer to this configuration as the \textit{baseline network}. In this configuration, a drone is placed at the centroid location of its users. We use a $k$-means clustering algorithm to find the centroid location for all the drones \cite{lyu2016placement}. 

We consider that the available power at each drone is a random variable. The distribution of this random variable is referred to as the \textit{type} of the drone. Every drone knows its own type; however, concerning other drones, it only knows the set of possible types, i.e., a set of possible distributions. The possible types depend on the environment and the location of each drone so that it can be inferred from the historical data. Each drone maintains a belief about the types of other drones.
\subsection{Channel Model}
\label{seubsec:ChModel}
Let $\mathcal{N}_d$ and $\mathcal{Q}_d$ denote the set of users and channels of the drone $d \in \mathcal{D}$. Moreover, let user $n \in \mathcal{N}_d$ and drone $d \in \mathcal{D}$ be located at $(x,y,0)$ and $(x_d, y_d, h_d)$, respectively. Then the path-loss of the downlink communication from drone $d$ to user $n$ over the channel $q$ is given by \cite{zhang2018machine} 
\begin{equation} 
\label{eq:1}
{ L_{dn,q}[dB] = 20 \log \left(\frac{4 \pi f_c d_{dn}(x,y)}{c}\right)  + \zeta_{dn} + 10 \log (\Omega_{dn})},
\end{equation}
where $d_{dn}(x,y)=\sqrt{(x_d-x)^2+(y_d-y)^2+(h_d)^2}$ is the distance between drone $d$ and user $n$. Moreover, $f_c$ and $c$ are the carrier frequency and the speed of light, respectively. Also, $\zeta_{dn}$ is the average loss due to the free-space propagation, which depends on the environment. If the wireless link between drone $d$ and user $n$ is LoS, then $\zeta_{dn}^{\text{LoS}}=N(\mu_{\text{LoS}}, \sigma^2_{\text{LoS}})$; otherwise, the non-line-of-sight (NLoS) link has a loss of $\zeta_{dn}^{\text{NLoS}}=N(\mu_{\text{NLoS}}, \sigma^2_{\text{NLoS}})$. Therefore, $\zeta_{dn}^{\text{LoS}}$ and $\zeta_{dn}^{\text{NLoS}}$ are shadow fading with normal distribution in dB scale for LoS and NLoS links. The expected value and variance of the shadow fading for LoS and NLoS links are $(\mu_{\text{LoS}}, \sigma^2_{\text{LoS}})$ and $(\mu_{\text{NLoS}}, \sigma^2_{\text{NLoS}})$, respectively. The variance depends on the elevation angle and type of the environment \cite{azari2016optimal}, i.e.
\begin{gather} 
\label{eq:bc}
{ \sigma_{\text{LoS}}(\theta_{dn}) = k_1 \exp(-k_2 \theta_{dn})} \\
{ \sigma_{\text{NLoS}}(\theta_{dn}) = g_1 \exp(-g_2 \theta_{dn})},
\end{gather}
where $\theta_{dn} = \sin^{-1}(h_d/d_{dn})$ is the elevation angle between drone $d$ and user $n$. The parameters $k_1$, $k_2$, $g_1$, and $g_2$ are constants that depend on the environment. Let $\Omega_{dn}$ be a random variable capturing the effects of the small-scale fading between drone $d$ and user $n$ with $\bar{\Omega}_{dn}=1$. The random variable $\Omega_{dn}$ follows a non-central chi-square probability distribution, given by \cite{azari2016optimal}
\begin{equation} 
\label{eq:4}
{ f_{\Omega_{dn}}(\rho) = \frac{ (K+1) e^{-K} }{\bar{\Omega}_{dn} } e^{ \frac{ (K+1) \rho }{\bar{\Omega}_{dn} } } I_0\left ( 2 \sqrt{\frac{ (K+1) \rho }{\bar{\Omega}_{dn} }}  \right ) }.
\end{equation}
In (\ref{eq:4}), $\rho \geq 0$, and $I_0(\cdot)$ is the zero-order modified Bessel function of the first kind. Moreover, $K$ is the Rician factor that corresponds to the ratio of the power of the LoS component and the power of the multipath components. For $K = 0$, the Rician model reduces to a Rayleigh fading distribution. Hence, in general a small value of $K$ represents that multipath component is dominant whereas a large value corresponds to a strong LoS between the drone and user. Thus, $\Omega_{dn}^{\text{LoS}} = f_{\Omega_{dn}}(\rho)$ and $\Omega_{dn}^{\text{NLoS}} = f_{\Omega_{dn}}(\rho)|_{K=0}$.
The Rician factor $K$ depends on some factors out of which the elevation angle $\theta$ between the drone and user plays the dominant role. We consider \cite{azari2016optimal}
\begin{equation} 
\label{eq:5}
{ K = \psi(\theta) = a e^ {b \theta}},
\end{equation}
where $a$ and $b$ are some constants whose values depend on the system parameters and the environment. The constants $a$ and
$b$ are determined as
\begin{equation} 
\label{eq:6}
{ a = k_0, \ b = \frac{2}{\pi} \left( \ln{  \frac{k_{\frac{\pi}{2}}}{k_0} } \right) },
\end{equation}
where $k_0$ and $k_{\frac{\pi}{2}}$ could be determined from measurements in a concrete scenario \cite{azari2016optimal}.
Let $p_{dn}^{\text{LoS}}$ be the probability of existence of the LoS link between a drone $d$ and user $n$, given by \cite{zhang2018machine}
\begin{equation} 
\label{eq:7}
{ p_{dn}^{\text{LoS}} = \alpha \left(\frac{180}{\pi}\theta_{dn} - \theta_{o} \right)^\gamma},
\end{equation}
where $\alpha$ and $\gamma$ are constant values reflecting the environment impact. Moreover, $\theta_o$ is the minimum angle between a user and a drone. Then $p_{dn}^{\text{NLoS}} = 1 - p_{dn}^{\text{LoS}}$ is the probability of having a NLoS link. The average path loss from drone $d$ to user $n$ is
\begin{equation} 
\label{eq:82}
{ \bar{L}_{dn,q}= p_{dn}^{\text{LoS}} L_{dn,q}^{\text{LoS}} + p_{dn}^{\text{NLoS}} L_{dn,q}^{\text{NLoS}}}.
\end{equation}
The signal-to-interference-plus-noise ratio (SINR) from drone $d$ to a user $n$ over channel $q$ yields
\begin{equation} 
\label{eq:SINR}
{ \text{SINR}_{dn,q} = \frac{p_{n} G/ \bar{L}_{dn,q} }{B_w(N_0 + I_0) }},
\end{equation}
where $p_{n}$ is the average transmission power of drone $d$ towards user $n$, $N_0$ and $I_0$ are noise power and interference power spectral density (e.g. due to co-channel interference from other networks), respectively, $B_w$ is the channel bandwidth, and $G$ is the antenna gain. The downlink transmission rate is then given by
\begin{equation} 
\label{eq:user_rate_eqn}
{R_{dn} (q, p_{n}) = B_w \log(1 + \text{SINR}_{dn,q})}.
\end{equation}
%
\section{Rate Maximization in Drone Networks: Problem formulation}
\label{subsec:Optimization}
\subsection{Optimization Problem Formulation}
Primarily, each drone serves its assigned users independently. However, this might not be optimal in several scenarios, as the users associated with one drone might have better channels (e.g. LoS link) to some other drone. Moreover, the amount of available power is not identical for all drones and also changes over time. Consequently, promoting cooperation among drones improves network performance significantly. For intelligent cooperation, the drones can be divided into clusters. Inside each cluster, the drones share their radio resources including spectrum and energy (e.g. energy sharing among drones can be possible through wireless power transfer \cite{Zhang19:WPT}). They also cooperate in serving users, irrespective of the initial association. In brief, by shuffling users and pooling the resources, the drones in each cluster improve the aggregate performance in terms of the transmission rate. To formalize the problem, we note the following:
\begin{enumerate}
\item The drones are clustered into disjoint groups. Let $\mathcal{W}$ with cardinality $W$ be the set of all possible partition structures. For any partition structure $w \in \mathcal{W}$, we have $w=\{C_1,\cdots,C_k,\cdots,C_l\}$, where $C_k \cap C_{k'}=\phi$ and $\bigcup_{k=1}^{l} C_k=\mathcal{D}$. Let $C_k \in w$ denote an arbitrary coalition. Moreover, $\mathcal{D}^{C_k}$ is the set of drones in $C_k$. Also, $\mathcal{N}^{C_k}$ and $\mathcal{Q}^{C_k}$ are the set of channels and users that the drones in $C_k$ share. 
\item Inside each cluster $C_k$, any user $n \in \mathcal{N}^{C_k}$ shall be served by one drone, say $d \in \mathcal{D}^{C_k}$. Each drone can serve multiple users. Matrix $\textbf{Y}^{C_k}$ of dimension $D^{C_k} \times N^{C_k}$ represents the user assignment. Therefore, in coalition structure $w$ with $l$ coalitions, there exists $l$ such assignment matrices. We gather these as $\textbf{Y}_w=\{\textbf{Y}^{C_1}, \dots, \textbf{Y}^{C_k}, \dots, \textbf{Y}^{C_l} \}$. The element $y_{dn}$ of the matrix $\textbf{Y}^{C_k}$ is defined as follows:
\begin{equation} 
\label{eq:user_assign}
y_{dn}=\begin{cases}
1, & \text{if user $n$ and drone $d$ are associated;} \\
0, & \text{otherwise.}
\end{cases}
\end{equation}
\item In addition to user assignment, each channel $q \in \mathcal{Q}^{C_k}$ is allocated to one user $n \in \mathcal{N}^{C_k}$. Matrix $\textbf{X}^{C_k}$ with dimension $Q^{C_k} \times N^{C_k}$ denotes the channel assignment. The set of channel allocation matrices for a partition structure $w$ is $\textbf{X}_w = \{\textbf{X}^{C_1}, \dots, \textbf{X}^{C_k}, \dots, \textbf{X}^{C_l} \}$. Each element $x_{qn}$ of the matrix $\textbf{X}^{C_k}$ is defined as
\begin{equation} 
\label{eq:user_assign2}
x_{qn}=\begin{cases}
1, & \text{if channel $q$ is allocated to user $n$;} \\
0, & \text{otherwise.}
\end{cases}
\end{equation}
\item Inside each cluster, the drones optimize the power allocation. Let $P_d$ denote the available power at drone $d$. The total power of the group $C_k$ then yields $P^{C_k} = \sum_{d = 1}^{\mathcal{D}^{C_k}}P_d$, where we neglect the loss that might occur due to energy sharing (e.g. via wireless power transfer). The power $P^{C_k}$ is then allocated to the users $\mathcal{N}^{C_k}$. Let $\textbf{p}^{C_k} = [p_1, \dots, p_n, \dots, p_{\mathcal{N}^{C_k}} ]$ be the allocation vector, where $\sum_{n = 1}^{\mathcal{N}^{C_k}}p_n \le P^{C_k}$.
\end{enumerate}
Let $R_{dn}(q,p_n)$ be the transmission rate provided by drone $d \in \mathcal{D}^{C_k}$ to user $n \in \mathcal{N}^{C_k}$ over channel $q \in \mathcal{Q}^{C_k}$ with power $p_n$. Based on the discussion above, the objective is to select the partition $w \in \mathcal{W}$, the user association matrices $\textbf{Y}_w$, the channel allocation matrices $\textbf{X}_w$, and the power allocation $\textbf{p}^{C_k}$ in each cluster $C_k \in w$ to maximize the total transmission rate. Formally, for all $C_k \in w $, $n \in \mathcal{N}^{C_k}$, $q \in \mathcal{Q}^{C_k}$, $d \in \mathcal{D}^{C_k}$, 
\begin{subequations} 
\begin{align}
& \underset{w,\textbf{Y}_w,\textbf{X}_w,\textbf{p}^{C_k}}{\text{maximize}}\sum_{C_k \in w} \sum_{n \in \mathcal{N}^{C_k}} R_{dn}(q,p_n) x_{qn} y_{dn} \\ 
&  \text{s.t.} \hspace{0.6cm} \sum_{d=1}^{\mathcal{D}^{C_k}} y_{dn} = 1, \quad y_{dn} \in \{0,1 \}  \label{subeq:main_user_assign_constaint} \\ 
& \hspace{1cm}\sum_{q=1}^{ \mathcal{Q}^{C_k}}\sum_{n=1}^{\mathcal{N}^{C_k}}x_{qn}=1, \quad x_{dn} \in \{0,1 \} \label{subeq:main_channel_assign_constaint1} \\
& \hspace{1cm} \sum_{n=1}^{\mathcal{N}^{C_k}} p_n \leq P^{C_k}, \quad p_{n}\geq 0 \label{subeq:power_constaint} \\ 
& \hspace{1cm} \sum_{n=1}^{\mathcal{N}^{C_k}}R_{dn}(q,p_n) y_{dn}>R_d^{(B)}. \label{subeq:drone_constaint}\\
\nonumber
\end{align}
\end{subequations}
Constraints \eqref{subeq:main_user_assign_constaint} and \eqref{subeq:main_channel_assign_constaint1}, respectively, ensure that any user is associated to only one drone and is allocated only one channel. Moreover, \eqref{subeq:power_constaint} guarantees positive allocated power for each user while the total allocated power does not exceed the available power. Constraint \eqref{subeq:drone_constaint} ensures that, for every drone $d$, the aggregate transmission rate of its associated users after cooperation, i.e. $\sum_{n=1}^{\mathcal{N}^{C_k}}R_{dn}y_{dn}$, is larger than the transmission rate provided by drone $d$ before cooperation (baseline configuration), denoted by $R_d^{(B)}$. \\
However, solving the formulated optimization problem is not feasible due to the following reasons: (i) Given limited information and uncertainty in the network parameters, the objective function might not be known; (ii) Mixed integer non-linear program formulation and  a combinatorial number of possible partitions (given by the Bell number) make the problem practically infeasible to solve. Therefore, we divide the optimization problem into three sub-problems: (i) clustering; 2) user-association and channel assignment; (iii) power control. Indeed, the drones first make coalition based on their limited information, and inside each coalition, they optimize the user association as well as the channel- and power allocation. 
\subsection{Clustering/Coalition Formation} 
\label{Partitioning}
Centralized partitioning requires full information and it is computationally expensive. Hence we allow drones to form coalitions by local decision-making. When forming coalitions, each drone acts rationally to maximize its performance. Therefore, cooperation under uncertainty becomes challenging, as the optimal coalition and action are not known in advance. To address this challenge, we take advantage of Bayesian coalition formation games. Then $w$ is a \textit{coalition structure} and each group $C_k \in w$ in the partition is a \textit{coalition}. In every coalition, the utility of each drone and the coalition depends on user- and channel assignment, as well as power allocation. \textit{Coalition formation will be discussed in detail in Section \ref{sebsec:BCFG} and Section \ref{solution_approach}}.
\begin{figure}[!ht]
\centering
\includegraphics[width =0.95 \columnwidth]{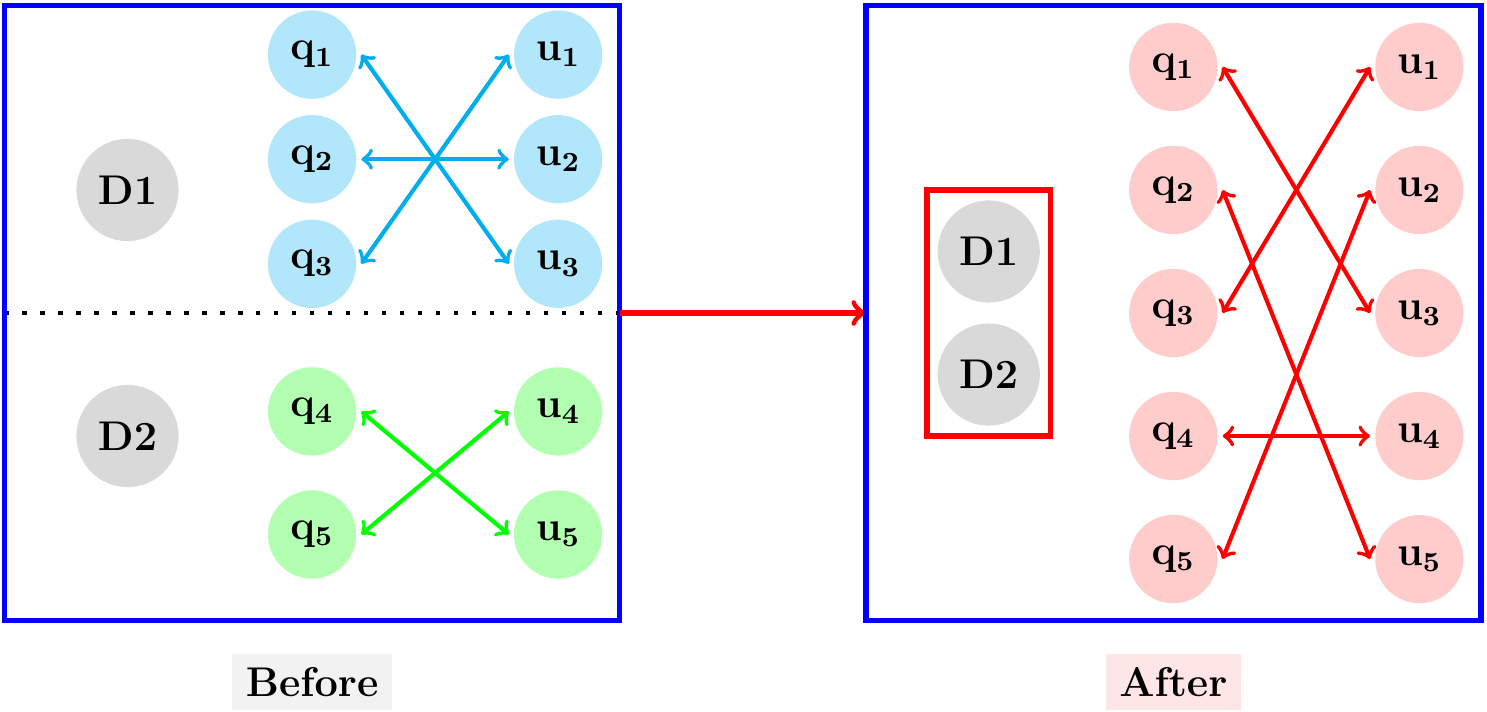}
\caption{Example of the channel assignment. The left figure shows the baseline configuration when drones $D1$ and $D2$ perform independently. By forming the coalition $\{D1,D2\}$, the drones share their spectrum resources, thereby enabling more channel assignment choices.} 
\label{fig:channel_association}
\end{figure}
\subsection{User and Channel Assignment} 
\label{joint_association}
For any coalition $C_{k}$, we formulate the user- and channel assignment problem as a weighted bipartite matching problem. We then use Hungarian algorithm \cite{kuhn1955hungarian} to solve the problem. 
\begin{definition}[Weighted Bipartite Matching] 
Let $G = (\mathcal{V}, \mathcal{E})$ be a weighted bipartite graph, where $\mathcal{V}$ consists of two set of vertices $\mathcal{V}_1$ and $\mathcal{V}_2$, such that $\mathcal{V} = \mathcal{V}_1 \cup \mathcal{V}_2$ and $\mathcal{V}_1 \cap \mathcal{V}_2 = \varnothing$. Moreover, $\mathcal{E} \subseteq \mathcal{V}_1 \times \mathcal{V}_2$ is the set of edges connecting the vertices. Let $e(i,j)$ denotes the edge between a vertex $i \in \mathcal{V}_1$ and $j \in \mathcal{V}_2$, and $w_{ij}$ represents the weight of the edge $e(i,j)$. The weights are represented by the graph matrix $\textbf{W}=[w_{ij}]$ of dimension $V_1 \times V_2$. \\
\textit{Matching}: A matching is the subset of edges $\mathcal{M} \subseteq \mathcal{E}$ such that for $e, e' \in M$, there is no vertex $v$ on which both the edges $e$ and $e'$ incident.\\  
\textit{Maximum (minimum) matching}: Let $W_{\mathcal{M}}$ represent the total weight of the selected edges in the matching $\mathcal{M}$. A matching $\mathcal{M}$ is maximum (minimum) if for any other matching $\mathcal{M}'$, we have $W_{\mathcal{M}} \geq W_{\mathcal{M}'}$ ($W_{\mathcal{M}} \leq W_{\mathcal{M}'}$).\\ 
\end{definition}
Concerning the user- and channel assignment problem in any coalition $C_k$, $\mathcal{V}_1$ and $\mathcal{V}_2$ represent the set of channels $\mathcal{Q}^{C_k}$ and the set of users $\mathcal{N}^{C_k}$, respectively. The weight of the an edge between a channel $q \in \mathcal{Q}^{C_k}$ and a user $n \in \mathcal{N}^{C_k}$, denoted by $w_{qn}$, is equal to the inverse average path-loss $1/\bar{L}_{dn,q}$. We then use the maximum bipartite matching for the assignment, i.e. we maximize the aggregate channel gains while associating users with drones/channels. For a given power allocation, such an assignment is equivalent to maximizing the overall transmission rate, as logarithmic function is monotone increasing. 
Note that the users are not explicitly assigned to the drones, rather implicitly by including fading- and shadowing effects in the overall channel gains of the users. 
\textbf{Fig. \ref{fig:channel_association}} shows an example of user- and channel allocation.
\subsection{Power Control} 
\label{power_control}
After channel assignment, the drones perform power allocation, e.g. using water-filling algorithm \cite{Gallager93}, given constraint (\ref{subeq:power_constaint}). By the consensus among the drones in every coalition, any other power allocation strategy can be used without affecting the procedure of coalition formation. 
\section{Bayesian Coalition Formation Game}
\label{sebsec:BCFG}
\subsection{Formulation of Bayesian Coalition Formation Game}
We formulate the problem of distributed cooperation among drones under uncertainty as a Bayesian coalition formation game (BCFG) \cite{chalkiadakis2012sequentially} with nontransferable utility (NTU). The game is the tuple $G: \big \langle \mathcal{D}, \mathcal{T},B, \mathcal{A}, \vec{\bar{\textbf{q}}}_{d \in \mathcal{D}} \rangle$, where 
\begin{itemize}
\item $\mathcal{D}=\{1,\dots,D \}$ is the set of  drones.
\item $\mathcal{T}$ is the set of {\em types}. The {available power} at each drone follows a distribution belonging to the \textit{type set} $\mathcal{T}=\{T_1,T_2, \dots,T_M \}$, where $M$ are the number of \textit{types}. The space of drones' joint type is then $\textbf{T} \in \otimes_{d=1}^{D}\{T_1,T_2,\dots,T_M \}$. Every drone $d \in \mathcal{D}$ knows its own type $t^d \in \mathcal{T}$, i.e. the distribution of its available power, but not those others. For each drone $d$, the type set of other drones yields $\textbf{T}^{-d} = \otimes_{x \in \mathcal{D} \backslash d}\{ T_1, T_2, \dots,T_M \}$. Note that the \textit{amount} of available power of any drone $d$ is the expected value of its type, i.e. $P_d = \mathbb{E}(t^d)$. The {\em type} of each drones remains fixed during the coalition formation process.
\item $B$ is the belief function. Let $\textbf{t}^{-d} \in \textbf{T}^{-d}$ be any vector from the set $\textbf{T}^{-d}$ where, $\textbf{t}^{-d} = [t^d_1, \dots,t^d_j \dots, t^d_{\mathcal{D}\backslash d}]$ and $t^{d}_j \in \mathcal{T}$ represents the type of the drone $j$ according to the drone $d$. Therefore, $B(\textbf{t}^{-d})$ is the joint belief of drone $d$ about others having type profile $\textbf{t}^{-d}$ as follows:
\begin{equation} \label{eq:joint_belief}
{B(\textbf{t}^{-d}) = \displaystyle\prod_{j \in \mathcal{D} \backslash d} \text{Pr} [t^d_j]},
\end{equation}
where $\text{Pr} [t^d_j]$ is the probability of the drone $d$ about drone $j$'s type. Similarly, the function $B(\textbf{t}_{C_k})$ indicates the marginal of $B$ over any coalition $C_k$ with members' types $\textbf{t}_{C_k} = \{ {t^d} \}_{d \in C_k} \in \textbf{T}_{C_k} = \otimes_{d \in C_k} \mathcal{T}$. Moreover, $B(\textbf{t}_{C_k}^{-d})$ indicates the joint belief of drone $d$ about the coalition members $j \in \{ C_k \backslash d \}$, where $\textbf{t}_{C_k}^{-d} \in \textbf{T}^{-d}_{C_k} = \otimes_{j \in \{C_k \backslash d\}} \mathcal{T}$. Further, $B^d(t^d)$ represents the drone $d$'s belief about its own {\em type}, which assign probability $1$ to its actual type and $0$ to all others.
\item $\mathcal{A}$ indicates the coalition actions. Indeed, each coalition $C_k$ has a set of coalition actions, denoted by $\mathcal{A}^{C_k}$. The coalition actions $\mathcal{A}^{C_k}$ is the set of all solutions for joint user association, channel assignments, and power allocations in coalition $C_k$. 
\item Based on the drone $d$'s belief, the available power of coalition $C_k$ is $P(\textbf{t}^{-d}_{C_k})=\sum_{j = 1}^{\mathcal{D}^{C_k}}P(t_j^d)$, where $P(t_j^d)$ is the power of the drone $j$ based on $d$'s belief $t_j^d$. The power is divided among users $\mathcal{N}^{C_k}$. Based on the {\em a priori} available channel information and the belief, we use weighted bipartite matching and the water-filling algorithm for channel selection and power allocation, respectively, as discussed in Section (\ref{joint_association}) and Section (\ref{power_control}). Let $R_{dn}(\textbf{t}_{C_k}^{-d}, a^{C_k})$ be the rate of the user $n$ connected to the drone $d$ in coalition $C_k$, calculated by drone $d$ based on the type $\textbf{t}_{C_k}^{-d}$ and action $a^{C_k} \in \mathcal{A}^{C_k}$. Let the selected channel and allocated power for user $n$ be $q$ and $p_n$, respectively, based on $a^{C_k}$ and $\textbf{t}_{C_k}^{-d}$. Thus, $R_{dn}(\textbf{t}_{C_k}^{-d}, a^{C_k}) = R_{dn}(q, p_n)$. The total transmission rate provided by done $d$ then yields $R_{d}(\textbf{t}_{C_k}^{-d}, a^{C_k}) = \sum_{n=1}^{\mathcal{N}^{C_k}} R_{dn}(\textbf{t}_{C_k}^{-d}, a^{C_k}) y_{dn}$. The expected payoff of drone $d$ is then given by
\begin{align} 
\label{eq:payoff}
{\bar{q}^d_d(B, a^{C_k}) = \sum_{\textbf{t}_{C_k}^{-d} \in \textbf{T}_{C_k}^{-d} } B(\textbf{t}_{C_k}^{-d})}
{R_d(\textbf{t}_{C_k}^{-d}, a^{C_k})}.
\end{align} 
Finally, the maximum achievable expected payoff of a drone $d$ in coalition $C_k$ is
\begin{equation} \label{eq:payoff1}
\bar{q}^d_d(B) = \max_{a^{C_k} \in \mathcal{A}^{C_k}} \bar{q}^d_d(B, a^{C_k}).
\end{equation}
Each drone aims at maximizing its own expected payoff by joining the best coalition.
\end{itemize}
\subsection{Belief Update Mechanism} 
\label{belief_update_mechanism}
At every iteration of coalition formation, the members of each coalition share the instantaneous information of the {available power (type)}. For updating its belief, each drone performs the following: (i) Each drone uses the maximum likelihood estimation (MLE) method to estimate the parameters of the distribution of the type; (ii) It uses Kullback-Leibler divergence (KL divergence) method to compare the closeness of the estimated parameters with the given set of types; (iii) Finally, each drone updates its belief about the types of other drones based on averaging. Details follow.
\subsubsection{The maximum likelihood estimation (MLE)}
\label{subsub:MLE}
In MLE, the objective function is the likelihood of the data $\textbf{X}$ given the model. The goal is to find the parameter $\theta$ that maximizes the evaluation function (the likelihood). Formally,
\begin{equation} 
\label{eq:15}
{\theta_{MLE} = \underset{\theta}{\text{argmax}}\  p(\textbf{X}| \theta)}.
\end{equation}
Let $\textbf{X} = [\textbf{x}_1, \textbf{x}_2, \dots , \textbf{x}_n]$ be the dataset consisting of $n$ samples of Gaussian process.\footnote{Here, we assume that the {available power} of each drone follows a normal distribution with unknown parameters. Adapting the entire analysis to any other distribution is straightforward.} The \textit{log likelihood} $\boldsymbol{\mathcal{LL}}$ of the data is given by
\begin{subequations} \label{eq:21}
\begin{align}
\boldsymbol{\mathcal{LL}} &= \log ( \mathcal{N}(\textbf{X}| \mu, \sigma^2)) = \sum_{n =1}^N \log ( \mathcal{N}(\textbf{x}_n| \mu, \sigma^2) \notag \\
&= \sum_{n=1}^{N}\log \left ( \frac{1}{\sqrt{2 \pi \sigma^2} } \exp^{\frac{1}{2}\left ( \frac{(\textbf{x}_n - \mu)^2}{\sigma^2} \right )} \right).
\end{align}
\end{subequations}
The best $\mu$ and $\sigma^2$ are obtained by taking  partial derivatives of above \textit{log likelihood} function $\boldsymbol{\mathcal{LL}}$ w.r.t. these parameters and equating them to zero. After solving the two equations, the estimated parameters are
\begin{equation} 
\label{eq:24}
{\mu_{MLE} = \frac{1}{N} \sum_{n=1}^{N} \textbf{x}_n; \quad \sigma^2_{MLE} = \frac{1}{N} \sum_{n=1}^{N} (\textbf{x}_n - \mu)^2}.
\end{equation}
%
\subsubsection{Kullback-Leibler divergence (KL divergence)} 
\label{subsub:KL}  
The KL divergence (also called relative entropy) is a measure of how one probability distribution is different from other. KL divergence of $0$ indicates that the two distributions are identical and $1$ indicates that they are completely different. Suppose $p$ and $q$ are the density of the normal random variables with mean $[\mu_1, \mu_2]$ and variance $[\sigma_1^2, \sigma_2^2]$. The KL distance from $q$ to $p$ is given by
\begin{equation} 
\label{eq:23}
{KL_{pq} = \log\left ( \frac{\sigma_2}{\sigma_1} \right ) + \frac{\sigma_1^2 + (\mu_1 - \mu_2)^2 }{2 \sigma_2^2} - \frac{1}{2}}.
\end{equation}
%
\subsubsection{Averaging}
\label{subsec:Avg}
The drones repeat the coalition formation process at $t = 1, 2, \dots $ until convergence. At every round, given the shared information, each drone re-estimates the parameters of the types' distributions (beliefs) using \eqref{eq:21}. Afterwards, it identifies the types using \eqref{eq:23}. It then calculates the average frequency of observation for each type, i.e. the number of times each type is observed so far over the total number of interactions.
%
\section{Distributed Algorithm for Bayesian Coalition Formation Game}
\label{solution_approach}
In this section, we propose a distributed method for coalition formation under uncertainty that consists of main parts: (i) Initialization of beliefs over the types, (ii) Coalition formation process based on the beliefs; and (iii) Updating the belief based on the local observations. \textbf{Algorithm \ref{Alg:algo1}} and \textbf{Algorithm \ref{Alg:algo2}}, respectively, summarize the overall procedure and the coalition formation process. 
\subsection{Algorithm 1: Repeated coalition formation under uncertainty} 
\label{algo1}
For the few first iterations, the drones form the grand coalition, where each drone shares the information about its instantaneous value of the type with other drones. The drones utilize this information to initialize the beliefs about each others' type by the belief updating mechanism discussed in section \ref{belief_update_mechanism}. If forming a grand coalition is not possible, the drones simply start from some arbitrary belief.

After initialization, at each round $t$, the controller draws a realization of some Bernoulli random variable $X$ with parameter $\epsilon$. If the outcome is $1$, the agents form the grand coalition; otherwise, they follow the distributed coalition formation process based on the \textit{best reply dynamics} discussed in \textbf{Algorithm \ref{Alg:algo2}}. The result is some coalition structure $w \in \mathcal{W}$. Inside each coalition, the drones share the instantaneous information. Given this information, each member of coalition $C_k$ updates the belief about its partners' types. Note that, forming the grand coalition with prob $\epsilon$ allows for enough information exchange so that the drones do not get stuck in forming some particular coalitions based on their beliefs. The process is repeated until there is no further change in the belief and a stable coalition structure is formed.

\begin{algorithm} 
\label{algo1FlowDiag}
\caption{Repeated coalition formation under uncertainty}
\label{Alg:algo1}
\begin{algorithmic}[1] 
\State \textit{\textbf{Initialization:}} Form the grand coalition for a few iterations. Initialize the belief $B$ given the observations and based on the mechanism \ref{belief_update_mechanism}. 
\Loop
\State A central controller draws a realization $x$ of a Bernoulli random variable $X$ with probability $\epsilon$. 	
\If { $X$ = 1 }
\State The drones form the grand coalition.
	\Else 
		\State Drones $\mathcal{D}$ engage in the coalition formation process using \textbf{Algorithm \ref{Alg:algo2}}. 
		\State \textbf{Algorithm \ref{Alg:algo2}} is run until Nash-stable/absorbing state $w^{\ast}$ is reached (i.e., when no drone benefits by leaving its coalition and join another coalition in the same coalition structure.)
		\EndIf
		\State Each drone broadcasts the instantaneous power information to the members of its coalition. 
		\State Based on this local information, each drone updates the belief about its coalition members using the belief update mechanism in Section \ref{belief_update_mechanism} . 
		\EndLoop 
	\end{algorithmic}
\end{algorithm}
\subsection{Algorithm 2: Distributed Coalition Formation Algorithm}
\label{algo2}
The dynamic coalition formation algorithm \cite{chalkiadakis2004bayesian},\cite{arnold2002dynamic} operates iteratively. First, the algorithm is initialized with any random feasible coalition structure. At every stage, a controller selects a drone $d$ with probability $1/D$, which is called the \textit{proposer}, to change its strategy.

The proposer $d$ observes the current coalition structure and decides for one of the following: (i) staying in the current coalition, (ii) joining another coalition in the same coalition structure, or (iii) forming the singleton coalition. Since the drones are rational, the proposer $d$ selects the coalition $S$ that results in the maximum expected payoff while believing that the affected members agree. Ties are broken simply at random. Thus, based on its belief, the proposer calculates its expected payoff in every coalition. It switches its coalition provided that its payoff is strictly higher and the expected payoff of other drones in the coalition $S$ where he joins, are at-least equal to their payoff before joining the proposer, i.e. they would agree that the proposer joins. The proposer then sends the request to the controller for joining its preferred coalition $S$. Upon receiving the proposal to join, all the members of coalition $S$ calculate their expected payoffs based on their beliefs conditioned that the proposer joins their coalition. They then compare the payoffs for before and after joining. Every drone in $S$ agrees that $d$ joins only if it does not result in a reduction in its expected payoff. Every drone in $S$ sends its positive or negative response to the controller, and the proposal is accepted only if all agree. If not allowed to join, the proposer repeats the same procedure with the next best coalition. This procedure is the \textit{best reply dynamics}.

The process continues until the drones achieve a stable coalition structure, also called the \textit{absorbing state}/Nash-stable coalition structure. In an absorbing state, no drone has an incentive to change its strategy, given the prevailing coalition structure. There can be multiple stable coalition structures (i.e. multiple absorbing states). The achieved absorbing state depends on the random sequence of the proposers. The best reply dynamics converges to one of the absorbing states with probability one as time tends to infinity \cite{chalkiadakis2004bayesian}.

\begin{figure}
\centering
\includegraphics[width = 0.75\columnwidth]{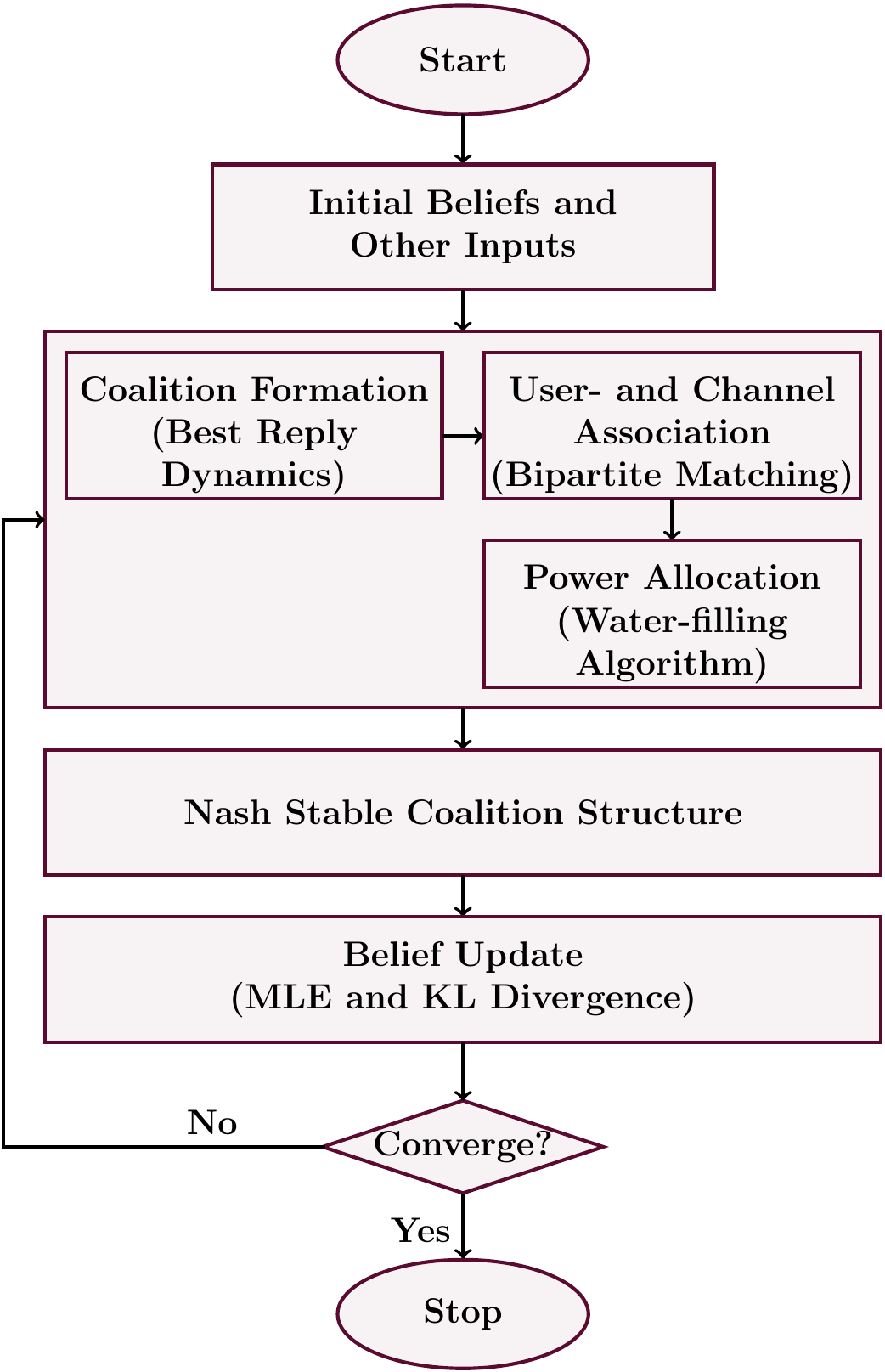}
\caption{Distributed coalition formation under uncertainty and resource allocation.} \label{fig:flow_chart}
\end{figure}

During the coalition formation process, the drones of each coalition share the information in a broadcast  control channel. Therefore, no pairwise interaction is required. As such, the signaling overhead remains low.
\textbf{Fig. \ref{fig:flow_chart}} shows the overall solution approach.
\begin{algorithm} 
\caption{Distributed coalition formation algorithm}
\label{Alg:algo2}
\begin{algorithmic}[1]
	\State Initialize $w(t)$.
	\Loop
	\State At time $t$, the controller selects a \textit{proposer} $d$ uniformly at random. Let $d$ belong to coalition $C_k$ of $w$. Upon being selected, it can decide to join some other coalition $S \in  w(t) \backslash C_k$, to remain in $C_{k}$, or to go singleton. 
	\State Based on its belief $B$, $d$ computes its expected payoff if joining any coalition $S$, $\bar{q}_d^d (S \cup \{d\})$, using (\ref{eq:payoff}). 
	\State Every drone $k \in S$ computes its expected payoff based on its belief for the case $d$ joins $S$ as $\bar{q}^k_k (S \cup \{d\})$. 
	\State Based on the computed expected payoffs, evaluate $C_k'=\arg \max_{S \in w(t) \backslash C_k} \bar{q}_d^d (S \cup \{ d \})$, such that it holds $\bar{q}_d^d (S \cup \{ d \}) > \bar{q}_{d}^d(C_k)$ and $\bar{q}^k_k (S \cup \{ d \}) \geq \bar{q}^k_k(S)$. 
	\If {$C_k'$ is not empty}
	\State 	drone $d$ leaves coalition $C_k$ and joins $C_k'$. Then the new state is
	\State $w (t + 1) = ((w(t) \backslash \{ C_k \}) \backslash \{ C_k' \} ) \cup \{ C_k' \cup \{ d \} \} \cup \{ C_k \backslash \{ d \} \} )$
	\Else 
	\State $w(t + 1)  = w(t)$
	\EndIf
	\State $t = t + 1$
	\EndLoop {\quad when the stable/absorbing state $w^{\ast}$ is reached.} 
	\end{algorithmic}
\end{algorithm}
\section{Analysis of the Bayesian Coalition Formation Game} 
\label{anaysis}
To analyze the stability of the formulated Bayesian coalition formation game, we use the concept of Nash stability/ absorbing states. We also develop a discrete-time Markov chain for the coalition formation algorithm that provides the Nash-stable solution.
\subsection{Convergence Analysis}
We start by stating the following definitions.
\begin{definition} 
Let $\mathcal{W} = \{ w_1,\dots,w_m,\dots, w_{W} \}$ be the set of states. A subset $F \subset \mathcal{W}$ is ergodic if, for $w \in F$ and $w' \not\in F$, the transition probability $\rho_{\omega,\omega'} = 0$ and no any other non-empty subset has this property. The singleton ergodic states are called absorbing states. A state $w$ is absorbing if $\rho_{w,w} = 1$. 
\end{definition}
Once the coalition formation process reaches the ergodic state, it remains there forever. 
\begin{definition}[Nash-stability/Absorbing state \cite{chalkiadakis2004bayesian} \cite{arnold2002dynamic}] Let drone $d$ belong to coalition $C_k(d)$ in state $w$. The state or a coalition structure $w = \{ C_1, \dots, C_k \dots, C_l \}$ is an absorbing (Nash-stable) state if for all $d \in \mathcal{D}$, for any proposal $S \cup \{ d \}$, either
\begin{itemize}
\item $\bar{q}_d^d(S \cup \{ d \}) \leq \bar{q}_d^d(C_k)$, or
\item $\exists j \in S $ such that $\bar{q}_j^j(S \cup \{d \}) < \bar{q}_j^j(S)$. 
\end{itemize}
The condition guarantees that \textit{either}
\begin{itemize}
\item The \textit{proposer} $d$ cannot improve its payoff by leaving its current coalition $C_k$ to join coalition $S$. For the tie $\bar{q}_d^d(S \cup \{ d \}) = \bar{q}_d^d(C_k)$, the drone remains in $C_{k}$ as it was already in effect and accepted by all the members, \textit{or} 
\item The proposal is denied by at-least one member of the coalition $S$, i.e. there is some agent $j \in S$ for which $q_j^j(S \cup \{d \}) < q_j^j(S)$. In this case, although the proposer has the incentive to join the coalition $S$, he cannot due to the denial.  
\end{itemize}
Thus, no drone $d \in \mathcal{D}$ leaves its current coalition to join another coalition in the same coalition structure. Also, no drone has an incentive to leave it current coalition to act alone. 
\end{definition}
The following proposition describes the convergence behavior of \textbf{Algorithm \ref{Alg:algo2}} \cite{chalkiadakis2004bayesian} \cite{arnold2002dynamic}.
\begin{proposition}
\textbf{Algorithm \ref{Alg:algo2}} converges to a Nash-stable solution or absorbing state \cite{chalkiadakis2004bayesian}.
\end{proposition}
\begin{definition}[Preference] 
Let a drone $d$ be the member of coalition $C_k$ in a coalition structure $w$. Then $S \succ_d C_k$ implies the strong preference of $d$ for coalition $S \in \{ w \backslash C_k \}$ over coalition $C_k$. The preference $S \succ_d C_k$ is valid if the following two conditions are satisfied:
\begin{itemize}
\item The expected payoff ($\bar{q}_d^d$) of drone $d$ is strictly higher when it joins coalition $S$ than when it remains a part of coalition $C_k$, i.e. $\bar{q}_d^d(S \cup \{ d \})>\bar{q}_d^d(C_k)$. 
\item No member of coalition $S$ believes that joining $d$ reduces its payoff, i.e. $\bar{q}_j^j(S \cup \{ d \}) \geq \bar{q}_j^j(S), \forall j \in S$. 
\end{itemize}
Similarly, $S \succeq_d C_k$ represents the weak preference of $d$ for $S$ over $C_k$. The weak preference means that the expected payoff of drone $d$ in coalition $S$ is at least equal to the expected payoff in its current coalition, i.e. $\bar{q}_d^d(S \cup \{ d \}) \geq \bar{q}_d^d(C_k)$. However, the conditions on the affected members remain same; that is, $\bar{q}_j^j(S \cup \{ d \}) \geq \bar{q}_j^j(S), \forall j \in S$. 
\end{definition}
\begin{definition}[Characteristic function]
Let $\vec{\bar{\textbf{q}}}^{\mathcal{D}} = [\bar{q}_1(\mathcal{D}),\dots, \bar{q}_d(\mathcal{D}), \dots \bar{q}_D(\mathcal{D})]$ be the expected payoff vector of all drones in grand coalition. Moreover,  $\vec{\bar{\textbf{q}}}^S = [\dots, \bar{q}_d(\mathcal{D}), \dots]$ denotes the expected payoff vector of the drones in any coalition $S \subset \mathcal{D}$. For coalition $S$, the characteristic function $V(\mathcal{D})$ is a set of feasible payoff vectors $\vec{\textbf{x}}^{S}$ of length $|S|$, where 
\begin{equation}
V(S) = \{ \vec{\textbf{x}}^{S} \in \mathbb{R}^{S} | \vec{\textbf{x}}^{S} \leq \vec{\bar{\textbf{q}}}^S \}.
\end{equation}
\end{definition}
\begin{definition}[Weak Bayesian core \cite{akkarajitsakul2011coalition}] 
The weak Bayesian core of an NTU game is defined as
\begin{align}
\mathcal{C} & = \{ \vec{\bar{\textbf{q}}}^{\mathcal{D}} \in V(\mathcal{D})| \forall S \subseteq \mathcal{D}, \nexists \ \vec{\bar{\textbf{q}}}^S \in V(S) \ \text{s.t.} \nonumber \\ 
& \qquad \qquad \qquad \qquad \qquad S \succeq_d \mathcal{D}, \forall d \in S  \}.
\end{align}
In words, in weak Bayesian core, the payoff of every drone in the grand coalition is such that no drone leaves the grand coalition as it is not better off in any other coalition, i.e. there is no blocking coalition.
\end{definition}
\begin{definition}[Strong Bayesian core \cite{akkarajitsakul2011coalition}]   
The strong Bayesian core of an NTU game is defined as
\begin{align}
\mathcal{C} & = \{ \vec{\bar{\textbf{q}}}^{\mathcal{D}} \in V(\mathcal{D})| \forall S \subseteq \mathcal{D}, \nexists \ \vec{\bar{\textbf{q}}}^S \in V(S) \
\text{s.t.} \nonumber \\ 
& \qquad  S \succeq_d \mathcal{D}, \forall d \in S \ \text{and} \ S \succeq_j^d \mathcal{D}, \forall j \in S, j \neq d \}.
\end{align}
The definition indicates that there exists a payoff profile in the grand coalition such that no drone has an incentive to leave the grand coalition. Also, every drone, based on its information about the expected payoff of others, believes that other drones are not better off if it leaves the grand coalition. The strong Bayesian core is the subset of the weak Bayesian core. Moreover, the conventional core is a special case of Bayesian core when there is no uncertainty, i.e. the drones know each others' types.
\end{definition}
\begin{remark}
The distributed coalition formation algorithm converges to a Nash stable solution but not necessarily to the Bayesian core.
For the proposed drone network, there may exist a blocking coalition for the grand coalition due to the individual rationality of the players/drones. However, Bayesian core exists if the grand coalition is the only stable coalition structure. If there are multiple Nash stable coalition structures and one of them is the grand coalition then the convergence to a particular Nash stable solution depends on the random sequence of proposers. The best case is to initialize the coalition structure with the grand coalition. If it is not stable, then the drones leave the grand coalition and form a stable coalition as a result of successive interactions. 
\end{remark}
\subsection{Discrete-Time Markov Chain-Based Analysis of Bayesian Coalition Formation Game}
\label{subsec:Discrete}
We formulate a finite state Markov chain to find the solution of coalition formation and to analyze the stable coalition structures \cite{chalkiadakis2004bayesian} \cite{arnold2002dynamic}. The state space of the Markov chain is the set of all coalition structures, i.e. $\mathcal{W}$. The transition matrix of the Markov chain is then $\textbf{W}$. Let drone $d$ belong to coalition $C_k(d)$ in the state $w$ at some time $t$. Moreover, $C_k'(d)=S \cup \{d\}$ is the coalition in state $w'$ at the next iteration $t+1$. Thus, the state $w'=((w \backslash \{ C_k\} \backslash \{ S \}) \cup (\{ C_k \backslash \{d \} \} \cup \{ S \cup \{ d \} \}) $. Transition from state $w$ to $w'$ depends on the decision of the proposer drone $d$ at time $t$. The transition probability $\rho_{w, w'}$, from state $w$ to $w'$, is given by
\begin{equation} 
\label{eq:8}
\rho_{w,w'}=\frac{1}{D}\sum_{d \in \mathcal{D}} \varphi_d(w'|w), 
\end{equation}
where $1/D$ is the probability that drone $d$ becomes a proposer to change its strategy. The term $\varphi_d$ is defined by best reply dynamics as follows:
\begin{itemize}
\item $\varphi_d=1/k_{w'}^d$, where $k_{w'}^d$ are the potential maximizers of the expected payoff of proposer $d$ at state $w$. The proposer joins one of these maximizers at random. The proposal to join the coalition $S$ by the proposer is feasible only if every members of the coalition $S$ accepts the proposal based on its private beliefs. Also, in the case $w \neq w'$, the $\varphi_d(w'|w)$ is non-zero for at most one $d$ in $\sum_{d \in \mathcal{D}} \varphi_d(w'|w)$. 
\item $\varphi_d=1$, if $C_k'(d)=S \cup \{d\}$ is the maximizer of the proposer $d$ but the proposal is rejected by at least one member of the affected coalition $S$ so that $w=w'$.
\item $\varphi_d=0$, otherwise. 
\end{itemize}
The stable coalitions are the absorbing states of the finite Markov chain. The stationary probability vector is $\vec{\boldsymbol{\pi}} = [\pi_{1} \dots \pi_{w} \dots \pi_{W}]^T$ where $\pi_{w}$ is the probability that the drones form the coalition structure $w$. Given that the transition matrix $\mathbf{W}$, the probability vector is the solution of 
\begin{equation}
{\vec{\boldsymbol{\pi}}^T \mathbf{W} = \vec{\boldsymbol{\pi}}^T}, \; \mbox{where } \vec{\boldsymbol{\pi}}^T \vec{\mathbf{1}} = 1.
\end{equation}
%
\section{Numerical Results} 
\label{results}
We consider an outdoor urban environment. The users that are initially assigned to each drone are randomly positioned inside the service area of $4 \text{km} \times 4 \text{km}$. Each drone is placed at the centroid of assigned users at a height of $1$ km above the ground. This configuration is the \textit{baseline} configuration. For a comprehensive evaluation, we implement four different settings concerning several network parameters. \textbf{Table \ref{table:simulation_settings}} summarizes the scenarios.  
\begin{table}[]
\centering
\caption{The parameters of simulated networks (each column represents a simulation setting)} \label{table:simulation_settings}
\begin{tabular}{|l|*{4}{c|}}\hline
\backslashbox{\textbf{Parameters}}{\textbf{Simulation} \\ \textbf{Settings}}
&\makebox[3em]{S1}&\makebox[3em]{S2}&\makebox[3em]{S3}
&\makebox[3em]{S4}\\\hline
Number of Drones ($D$) & 3 & 4 & 5 & 6 \\\hline
Number of Channels ($Q$) & 9 & 12 & 15 & 18 \\\hline
Number of Users ($N$) & 9 & 12 & 15 & 18 \\\hline
Users per Drone ($N_d$) & 3 & 3 & 3 & 3 \\\hline
Channels per Drone ($Q_d$) & 3 & 3 & 3 & 3 \\\hline
\end{tabular}
\end{table}

The drone-based communication operates over $2$ GHz carrier frequency ($f_c = 2$ GHz). The sum of noise and interference power spectral density $N_0+I_0$ is $-70$ dBm/Hz, antenna gain $G$ is $10$ dB and bandwidth $B_w$ is assumed to be 1 Hz. Without loss of generality, the bandwidth of each channel is $1$ Hz. The minimum angle $\theta_o$ is $15^{\circ}$. The small scale fading parameters are $k_0, k_{\frac{\pi}{2}} = [3, 30] $ dB  \cite{azari2016optimal}. \textbf{Table \ref{table:environment_parameters}} summarizes the parameters for different simulated environments.
For each network type, we generate $100$ topologies. We then simulate the coalition formation procedure for each topology $30$ times, as each topology can have multiple absorbing states, i.e. stable coalition structures. This means that each repetition of the coalition formation process at the same topology might converge into any of the possible absorbing states depending on the initial coalition structure and selection of the drones in the coalition formation process. It is also worth noting that for all topologies and simulation runs, the coalition formation process converges to a stable state in at most $50$ iterations. Convergence in such a short time is in particular interesting since, in theory, the convergence follows asymptotically.

\begin{table}[]
\centering
\caption{Environment parameters \cite{al2014modeling}} \label{table:environment_parameters}
\begin{tabular}{|c|c|c|c|c|}
\hline
\multirow{2}{*}{\textbf{Environment}} & \multicolumn{4}{c|}{\textbf{Parameters}}\\
\cline{2-5}
& $\alpha, \gamma$ & $k_1, k_2$ & $g_1, g_2$ & $\mu_{\text{LoS}}, \mu_{\text{NLoS}}$ \\
\hline
Urban & 0.6, 0.11 & 10.39, 0.05 & 29.06, 0.03 & 1, 20 \\\hline
Dense urban & 0.36, 0.21 & 8.96, 0.04 & 35.97, 0.04 & 1.6, 23 \\\hline
High-rise urban & 0.05, 0.61 & 7.37, 0.03 & 37.08, 0.03 & 2.3, 34 \\\hline
\end{tabular}
\end{table}

The proposed solution is compared with 
\begin{enumerate}
\item Baseline configuration, in which the drones do not cooperate. For each drone and its assigned users, the channel assignment and power allocation follow by bipartite matching and water-fill algorithm, respectively.
\item The solution using distributed best reply dynamics with complete information. That is, the drones have the full knowledge of channel quality as well as available power profiles. 
\item The socially optimal solution obtained using the exhaustive search, where the central controller has global information. The drones within each coalition share their channel and power. The coalition structure that has a maximum overall rate is selected as the optimal solution. Note that in this solution, achieving the highest aggregate performance has a higher priority compared to optimizing the performance for each drone. 
\end{enumerate}
%
\begin{figure}[!ht]
\centering
\includegraphics[width=0.85\columnwidth]{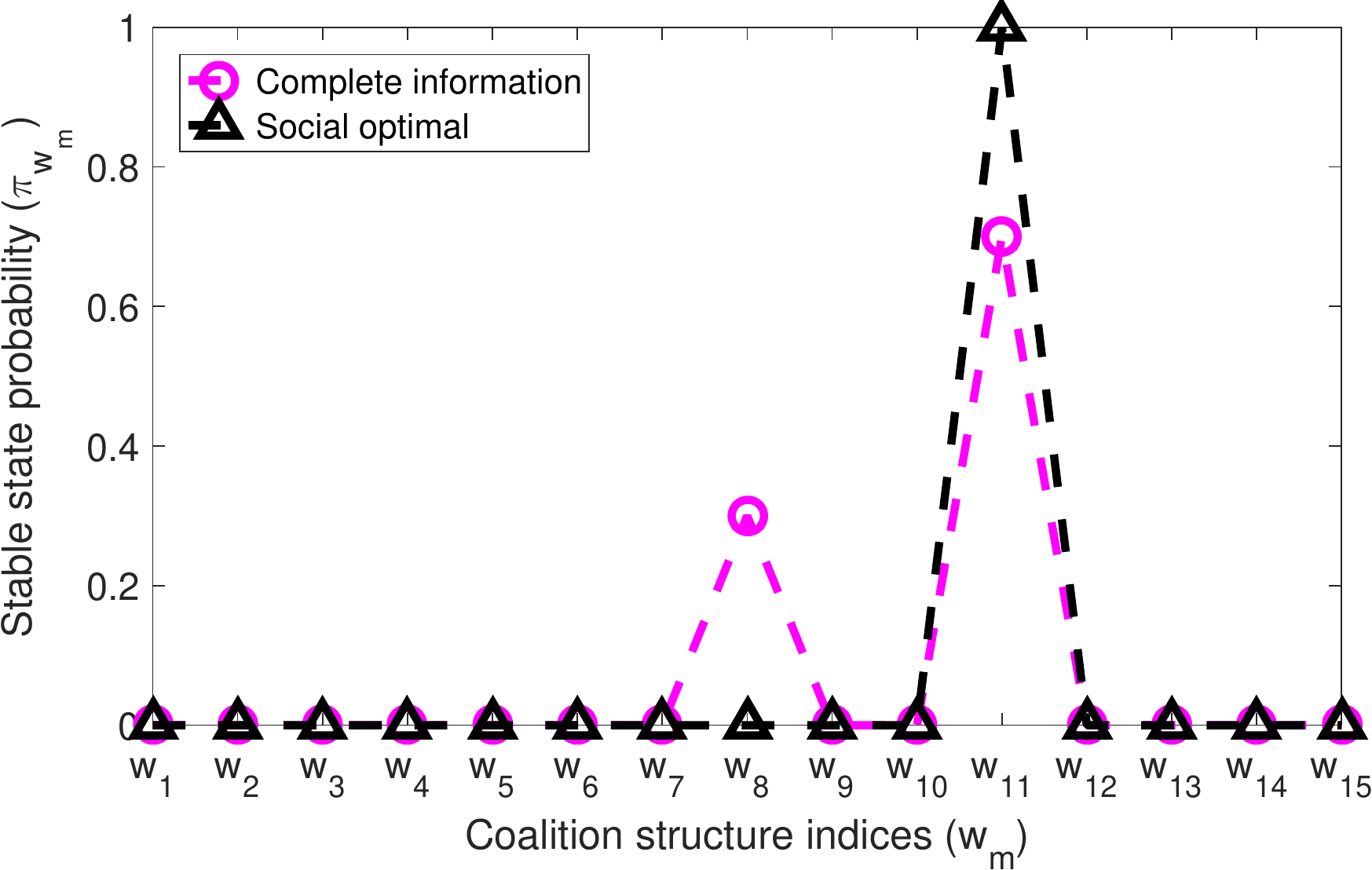}
\caption{Stationary probability of a network consisting of 4 drones and 3 users connected to each drone (Type 1: $\mu=12$, $\sigma=3$; Type 2: $\mu=18$, $\sigma=3$). }
\label{fig:stationary_prob_4drones}	
\end{figure}
\begin{figure}[!htb]
\centering
\includegraphics[width=0.85\columnwidth]{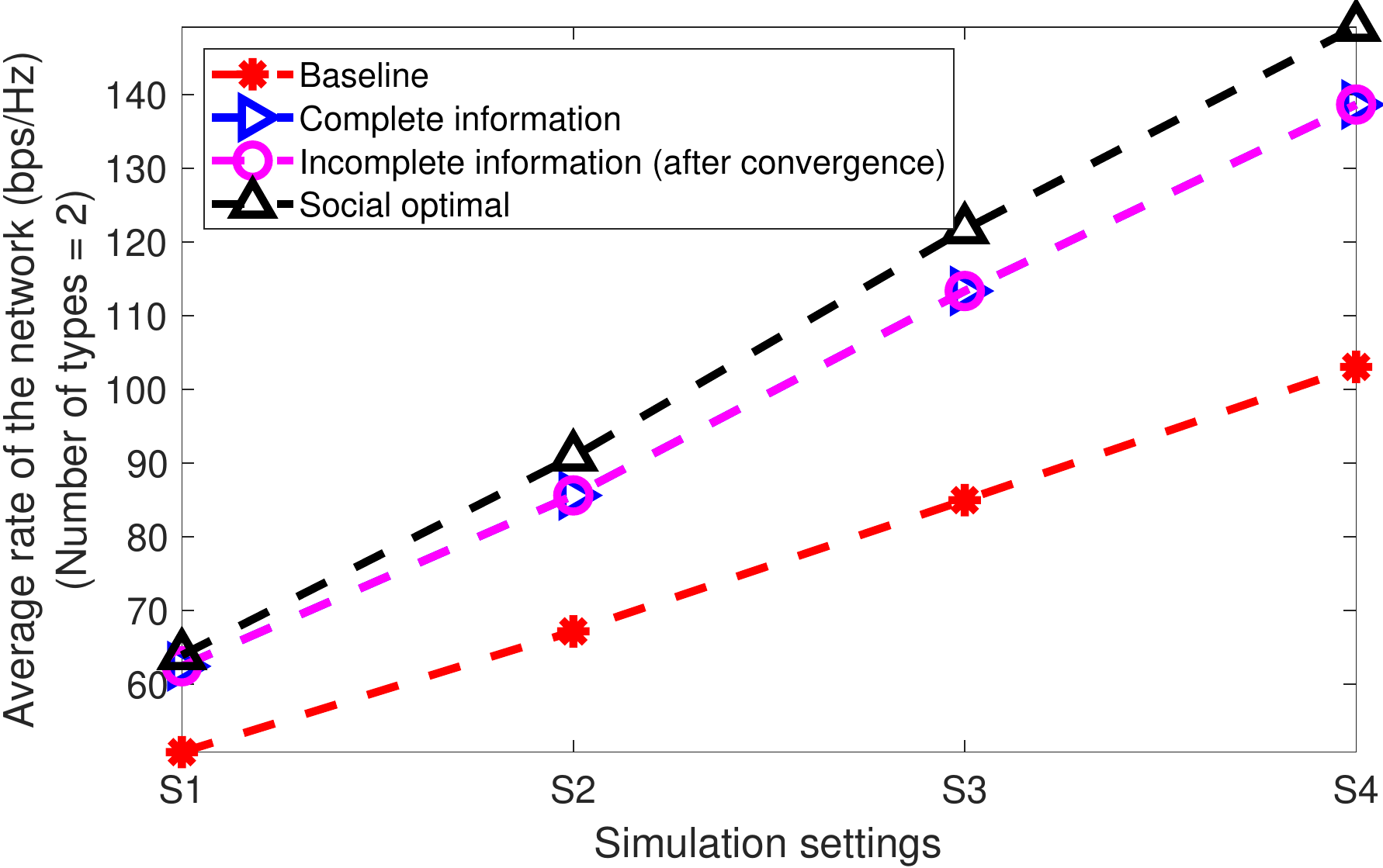}
\caption{Average transmission rate of the network for each simulation setting shown in Table~\ref{table:simulation_settings} for two types ($T_1: \mu=12,\sigma=3;\ T_2: \mu= 18, \sigma=3$). }
\label{fig:average_rate_case1_type2_power_low}	
\end{figure} 
\begin{figure}[!htb]
\centering
\includegraphics[width=0.85\columnwidth]{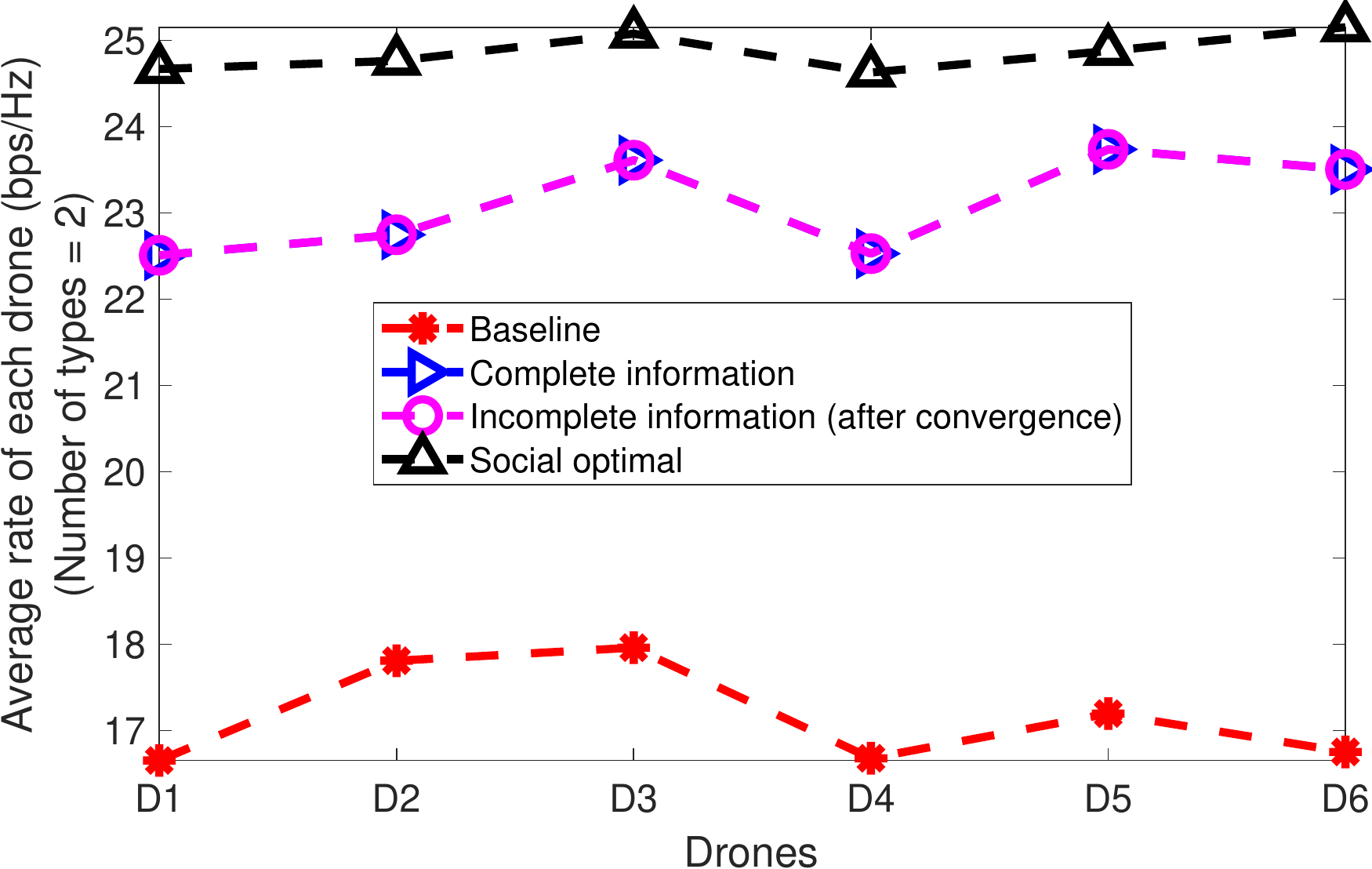}	\caption{Average transmission rate of each drone of the simulation setting with S4 (consists of $6$ drones) for two types as shown in Table~\ref{table:simulation_settings} ($T_1: \mu=12,\sigma=3;\ T_2:\mu= 18, \sigma=3$). }
\label{fig:average_individual_rate_case1_type2_power_low_6drones}	
\end{figure} 
\begin{figure}[!htb]
\centering
\includegraphics[width=0.85\columnwidth]{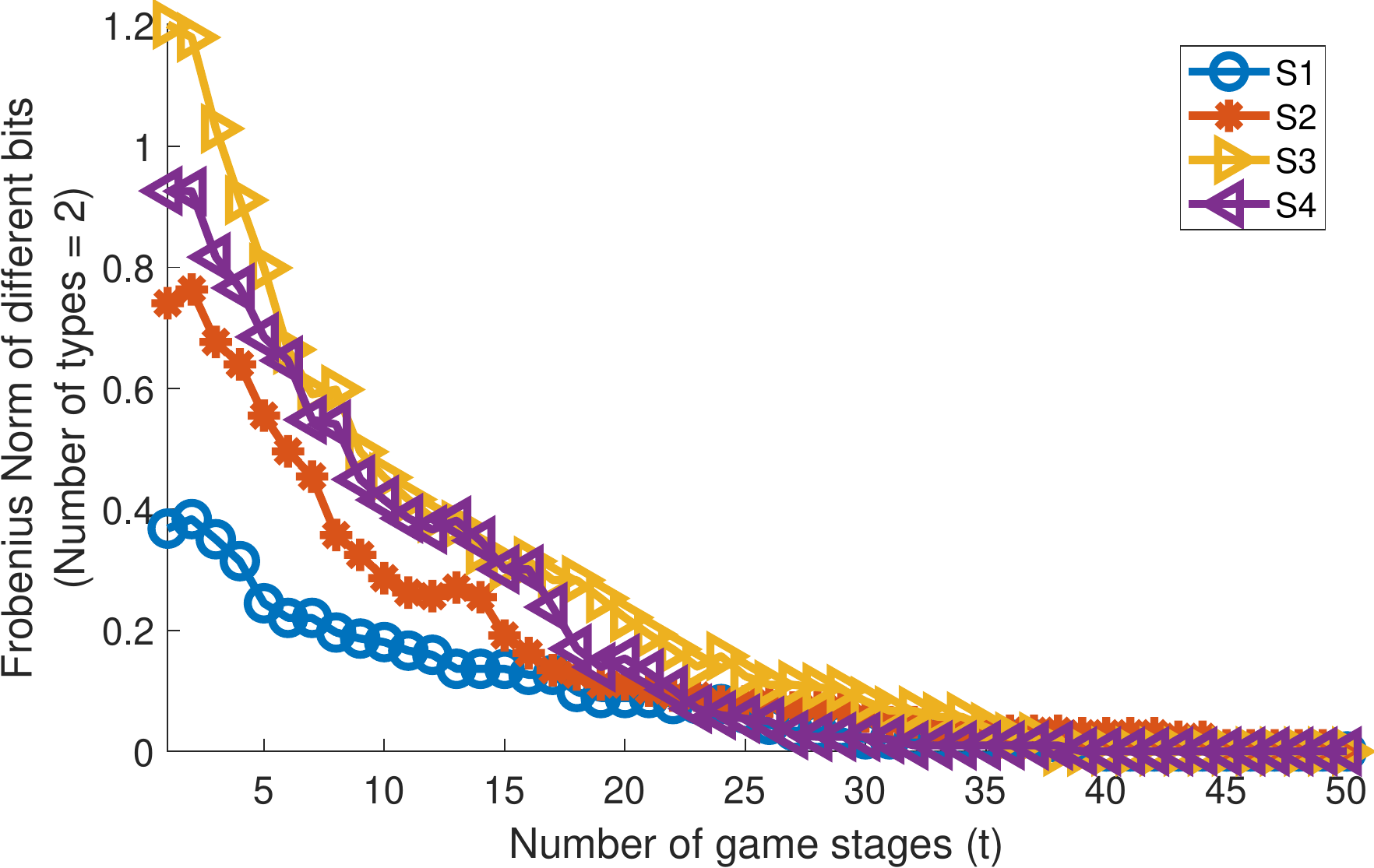}
\caption{Convergence for each simulation setting shown in Table~\ref{table:simulation_settings} for two types ($T_1: \mu=12,\sigma=3;\ T_2: \mu=18, \sigma=3$).}
\label{fig:convergence_two_types}	
\end{figure} 
For the distributed coalition formation process, for every topology, there are two ways to define the network's overall transmission rate:
\begin{itemize}
\item The maximum expected transmission rate for the network achieved by any stable coalition (the approach adopted in this work); 
\item The expected transmission rate of all the stable coalition structures. Formally,
\begin{equation} 
\label{average_expected_rate}
{\bar{q}_{total} = \sum_{d \in \mathcal{D}} \sum_{w = 1}^{W} \pi_{w} \bar{q}_d^d(C_k^d), \quad \text{for} \quad C_k^d \in w}.
\end{equation}
\end{itemize}
We present the stationary probabilities obtained from the Markov model in \textbf{Fig. \ref{fig:stationary_prob_4drones}}. We show the stationary probabilities of an arbitrary topology from $100$ random topologies of simulation setting $S4$ given in \textbf{Table \ref{table:simulation_settings}}. It consists of $4$ drones with $2$ types. Each drone has $3$ channels and $3$ users. The network of $4$ drones has a total of $15$ possible coalition structures $\{ w_1, \dots, w_{15} \}$, where $w_{8}$ = $\{ \{ d_1\}, \{ d_2, d_3, d_4 \} \}$ and $w_{11}$ = $\{ \{ d_1, d_2, d_3 \}, \{ d_4 \} \}$. For this specific topology, the best coalition structure is $w_{11}$ that provides the maximum sum rate for all drones. Thus, if drones seek to maximize the social welfare, the probability of forming $w_{11}$ is $1$. However, the best-reply dynamics with complete information yields the coalition structures $w_{11}$ and $w_8$ as absorbing states. Thus, $w_{11}$ and $w_8$ are formed $70 \%$ and $30 \%$ of the time, respectively, if the drones seek individual rationality ($\pi_{w_{11}} = 0.7, \pi_{w_{8}} = 0.3$).

\textbf{Fig. \ref{fig:average_rate_case1_type2_power_low}} shows the average total transmission rate of the network for four simulation settings given in \textbf{Table \ref{table:simulation_settings}}. We consider two {\em types} that are Gaussian random variables defined as $T_1: \mu= 12, \sigma=3;\ T_2: \mu= 18, \sigma=3$. We observe that the average transmission rate increases by moving from simulation setting $S1$ to $S4$. This happens since increasing the number of drones, users, and channels, improves the chances of experiencing good channel quality. 

\textbf{Fig. \ref{fig:average_individual_rate_case1_type2_power_low_6drones}} shows the performance improvement for individual drones for the simulation setting $S4$. 

\textbf{Fig. \ref{fig:convergence_two_types}} shows the learning of drones type over time. We quantify the progress of learning using the \textit{Frobenius norm}, as described in the following. Each drone updates its belief about the type of other drones after each iteration of coalition formation based on the local information. For a network of $D$ drones and $M$ types, every drone $d \in \mathcal{D}$ maintains a matrix $\textbf{T}$ of dimension $D \times D$ for each type, thus there are $M$ such matrices. After every iteration, any element $a_{ij}$ of the matrix for a given type is $1$ if the drone $i$ predicts that type for drone $j$ and $0$ otherwise. We do averaging after every iteration to obtain the probability of predicting the type by every drone for others in the network. For the evaluation of convergence, we convert the probability values to binary values. We then use the Frobenius norm to show the average number of bits that differ from true type bits. The Frobenius norm of a $m \times n$ matrix $\mathbf{A}$ is defined as 
\begin{equation} \label{Frobenius_Norm_eqn}
\norm[\bigg]{\mathbf{A}_{F}}=\sqrt{\sum_{i=1}^m \sum_{j=1}^n |a_{ij}|^2},
\end{equation}
where matrix $\mathbf{A}_F$ is difference of the matrix with true bits and the matrix obtained after belief update at every iteration. Frobenius norm $0$ represents the convergence to true value.  

\begin{figure}[t]
\centering
\includegraphics[width=0.85\columnwidth]{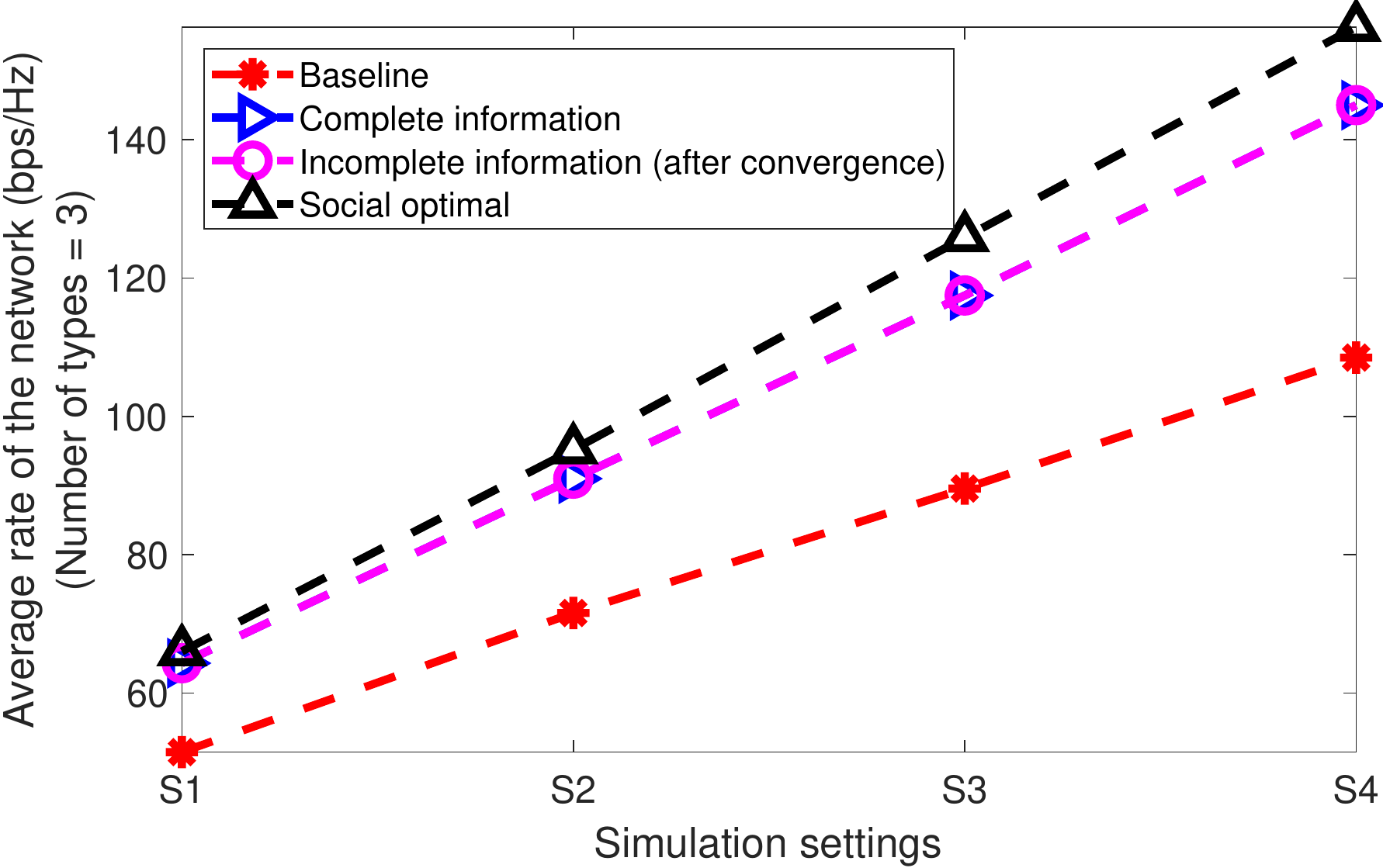}
\caption{Average transmission rate of the network for each simulation setting shown in Table~\ref{table:simulation_settings} for three types ($T_1: \mu=12, \sigma=3;\ T_2: \mu=18, \sigma= 3;\ T_3: \mu= 24, \sigma= 3$). }	\label{fig:average_rate_case2_type3_power_low}
\end{figure}

\begin{figure}[t]
\centering
\includegraphics[width=0.855\columnwidth]{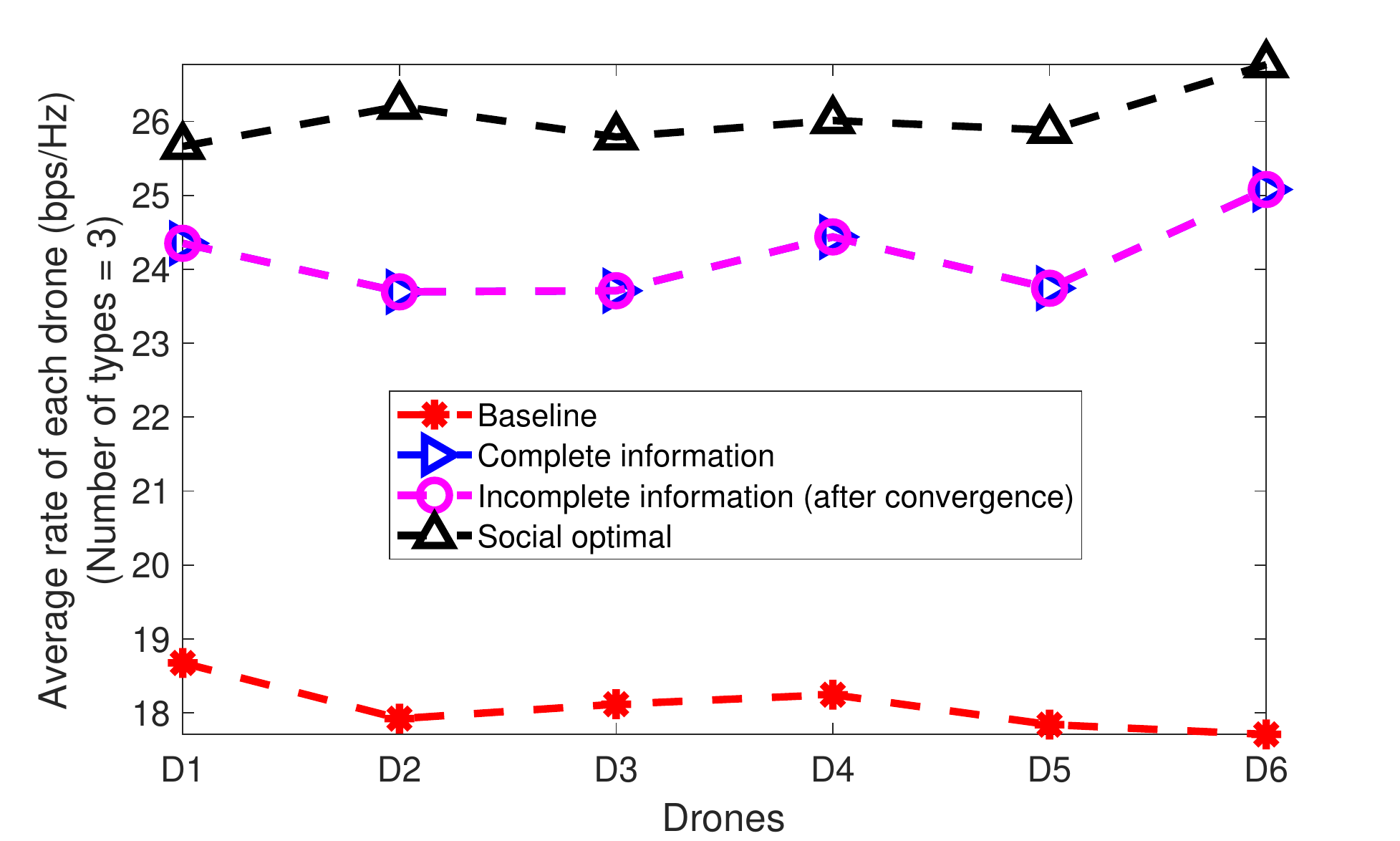}
\caption{Average transmission rate of each drone of the simulation setting with S4 (consists of $6$ drones) for three types as shown in Table~\ref{table:simulation_settings} ($T_1: \mu=12, \sigma=3;\ T_2: \mu= 18, \sigma= 3\ T_2: \mu= 24, \sigma= 3$). }
\label{fig:average_individual_rate_case2_type3_power_low_6drones}	
\end{figure}

\begin{figure}[t]
\centering
\includegraphics[width=0.85\columnwidth]{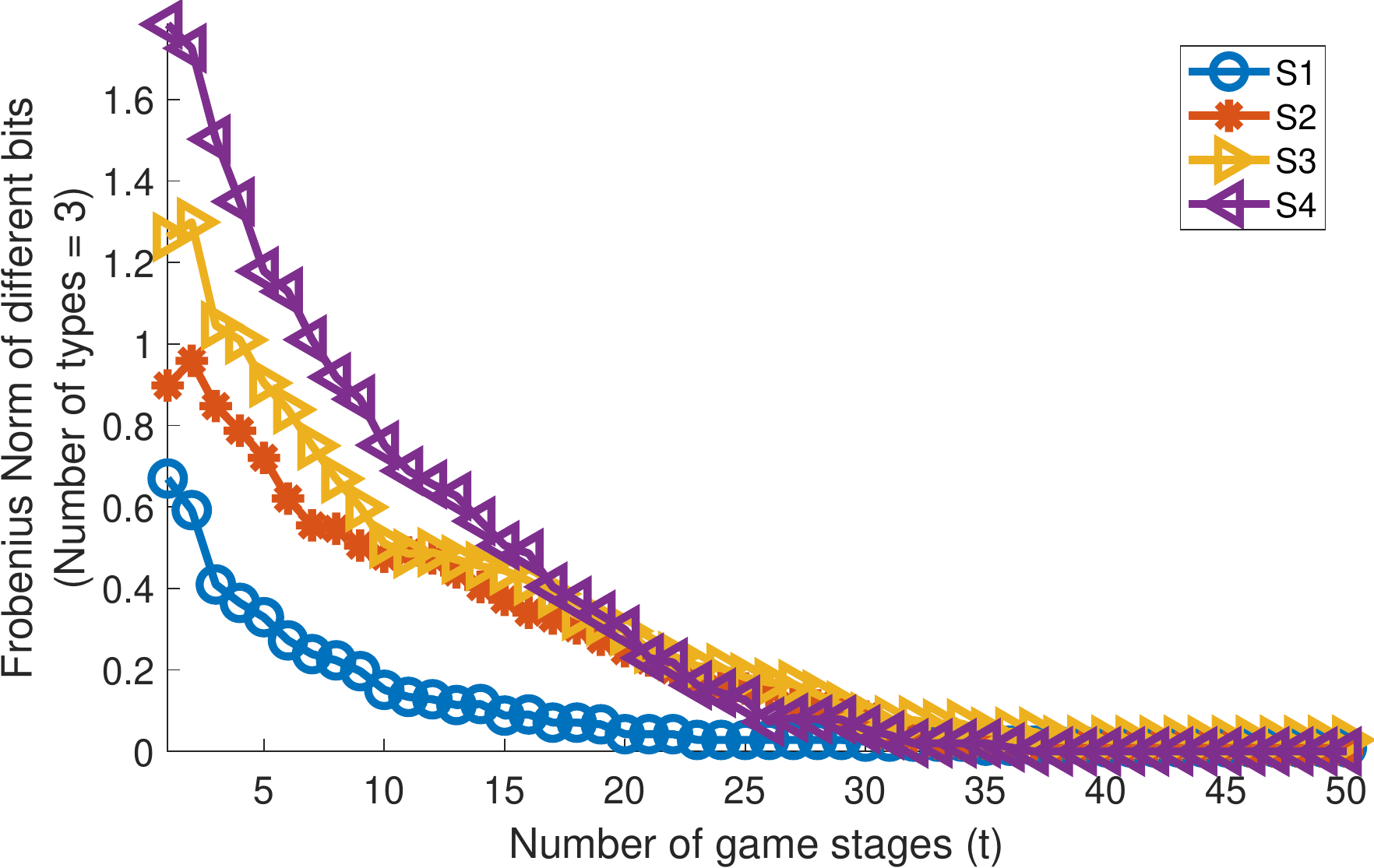}
\caption{Convergence for each simulation setting shown in Table~\ref{table:simulation_settings} for three types ($T_1: \mu=12, \sigma=3;\ T_2: \mu= 18, \sigma= 3;\ T_3: \mu= 24, \sigma= 3$). }
\label{fig:convergence_three_types}	
\end{figure}

Similarly, \textbf{Fig.  \ref{fig:average_rate_case2_type3_power_low}}, \textbf{Fig.  \ref{fig:average_individual_rate_case2_type3_power_low_6drones}}, and \textbf{Fig. \ref{fig:convergence_three_types}} show the performance for the same simulations setting given in Table \ref{table:simulation_settings} but for three Gaussian types as  $T_1: \mu=12, \sigma=3;\ T_2: \mu= 18, \sigma= 3;\ T_3: \mu= 24, \sigma= 3$. Here, the trend is similar to the scenario with two types.

\begin{figure}[!htb]
\centering
\includegraphics[width=0.85\columnwidth]{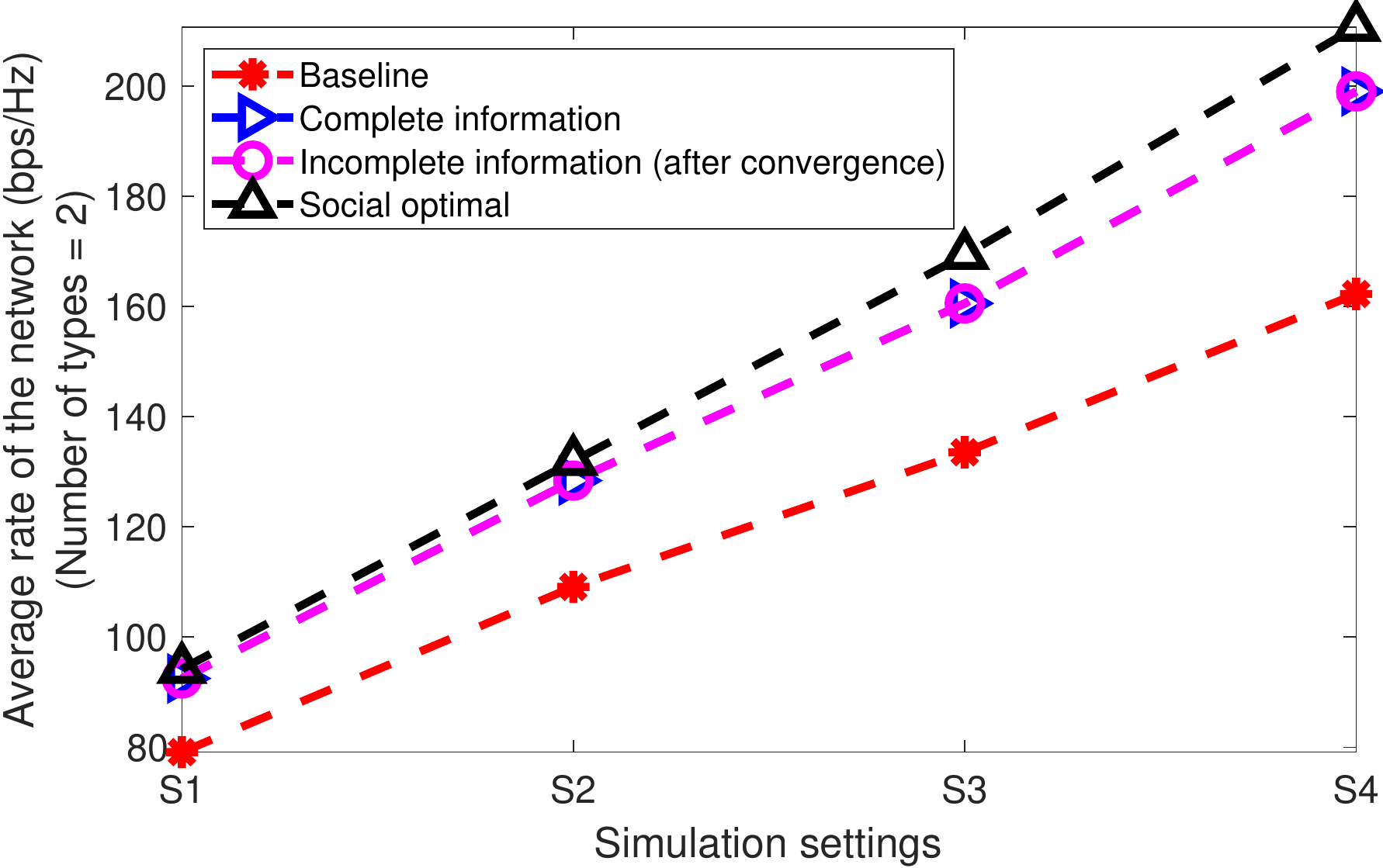}
\caption{Average transmission rate of the network for each simulation setting shown in Table~\ref{table:simulation_settings}, with the increase of the power level and two types ($T_1: \mu= 120, \sigma= 5;\ T_2: \mu=180, \sigma= 5$). }	
\label{fig:average_rate_case3_type2_power_high}	
\end{figure} 

In \textbf{Fig. \ref{fig:average_rate_case3_type2_power_high}}, we show the effect of increasing the power level. In this case, the Gaussian types are  $T_1: \mu=120, \sigma=5;\ T_2: \mu=180, \sigma= 5$. The figure shows that the network's transmission performance improves compared to the previous case of low power availability.  

\begin{figure}[!htb]
\centering
\includegraphics[width=0.85\columnwidth]{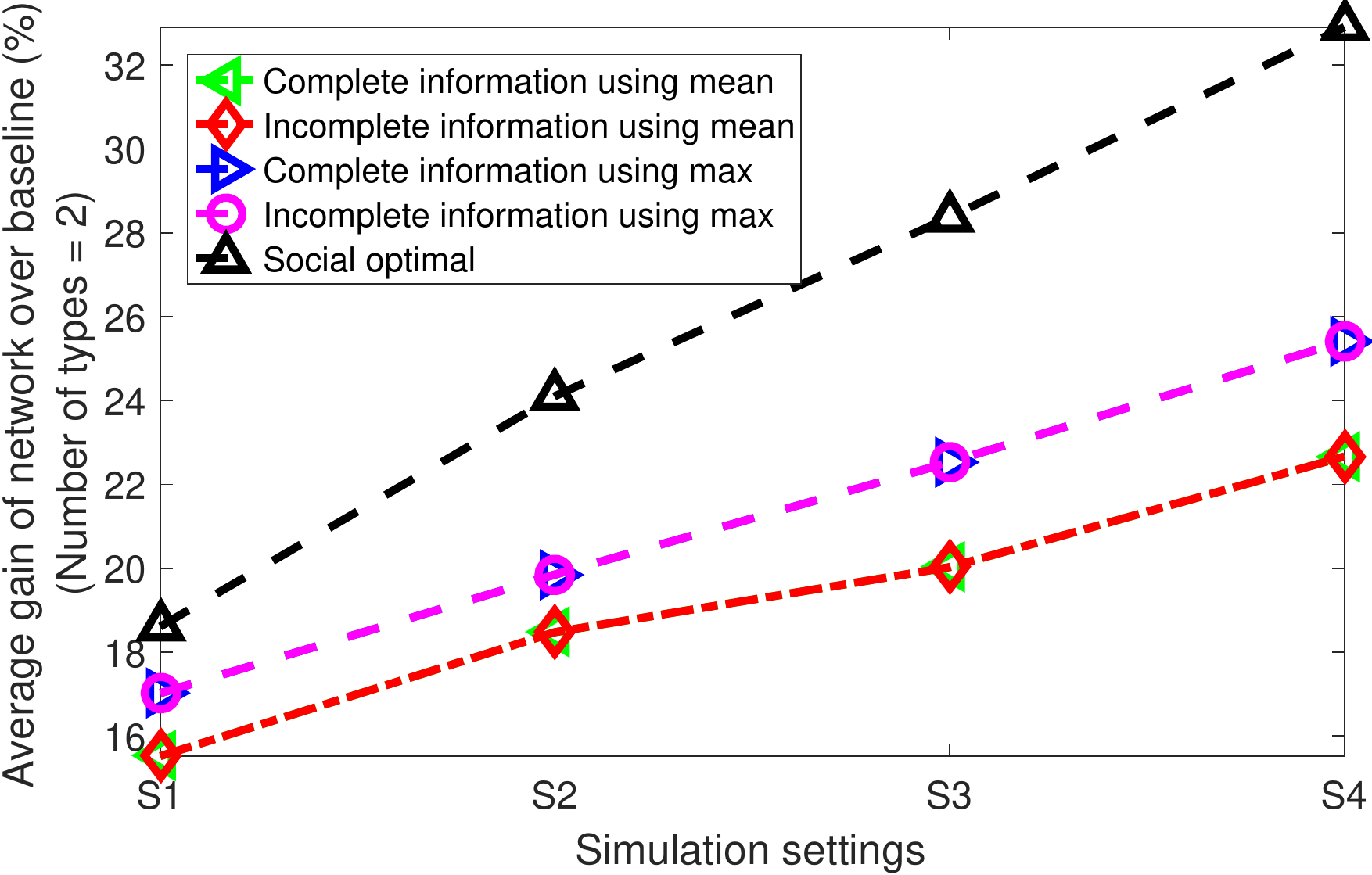}
\caption{Average transmission gain for each simulation setting shown in Table~\ref{table:simulation_settings} for two types ($T_1: \mu= 120, \sigma=5;\ T_2: \mu= 180, \sigma= 5$). } 
\label{fig:max_mean_average_gain_case3_type2_power_high}	
\end{figure}

In \textbf{Fig. \ref{fig:max_mean_average_gain_case3_type2_power_high}}, we show the comparison of the two different approaches of calculating the overall network transmission rate. We observe that by choosing the Nash-stable coalition structure having the maximum sum rate, the overall network's performance improves better compared to the approach of calculating the expected rate over all possible Nash-stable coalition structures, as given by (\ref{average_expected_rate}). 

\begin{figure}[!htb]
\centering
\includegraphics[width=1\columnwidth]{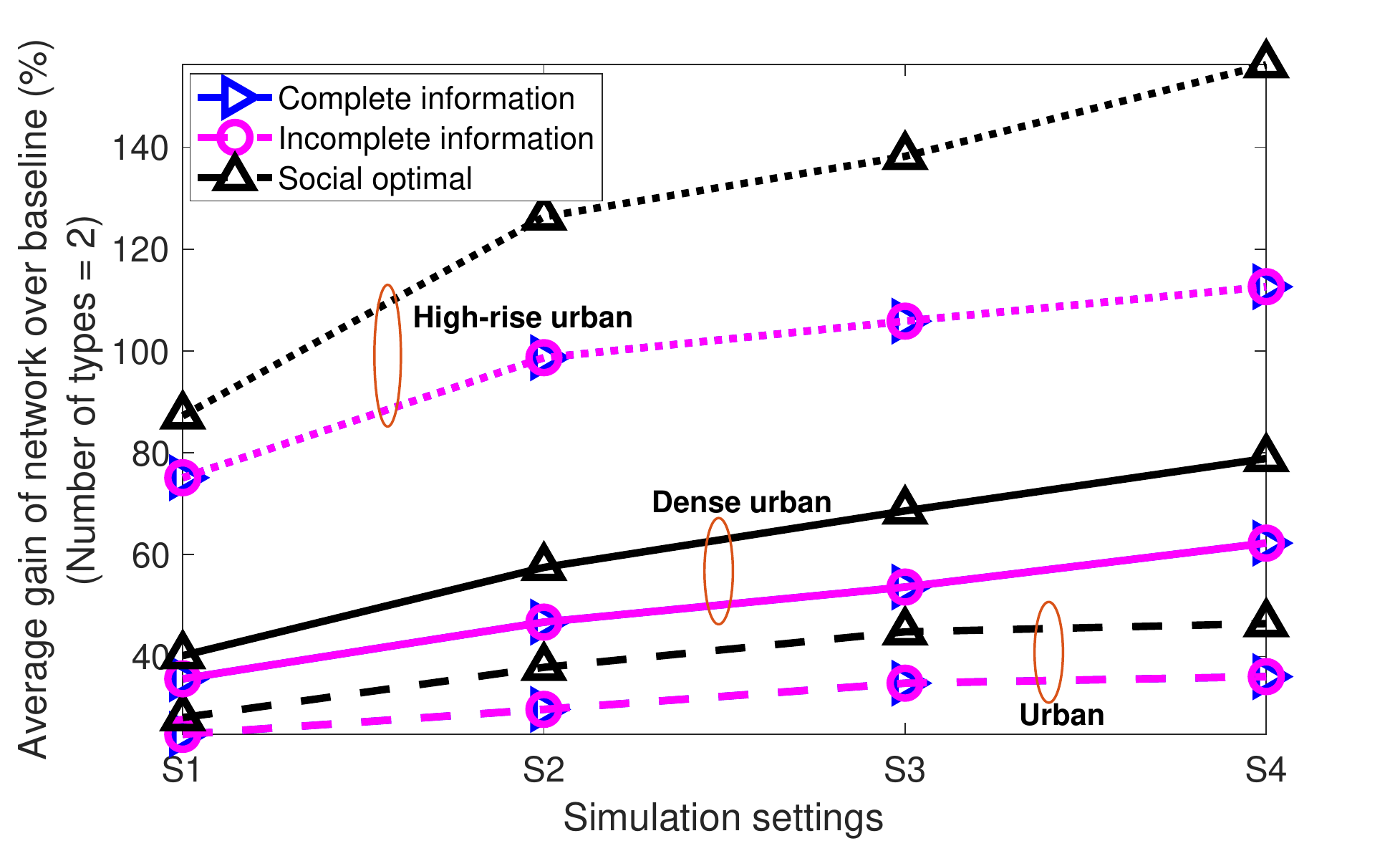}
\caption{Average transmission gain for various environments for two types ($T_1: \mu= 12, \sigma=3;\ T_2: \mu= 18, \sigma= 3$). } 
\label{fig:average_gain_type2_power_low_various_environments}	\end{figure}

\textbf{Fig. \ref{fig:average_gain_type2_power_low_various_environments}} shows the network's performance for various environments. We observe that compared to the dense-urban and urban environments, the performance of cooperation is better in high-rise urban environments. This is due to severe reflections, scattering, and shadowing in high-rise urban environment. In other words, leveraging cooperation among drones improves the network's performance substantially, as a high-quality LoS link towards a helping drone might be available to some users that have a low-quality channel towards the original serving drone. 

\begin{figure*}[!h]
\centering
\includegraphics[width=2.2\columnwidth]{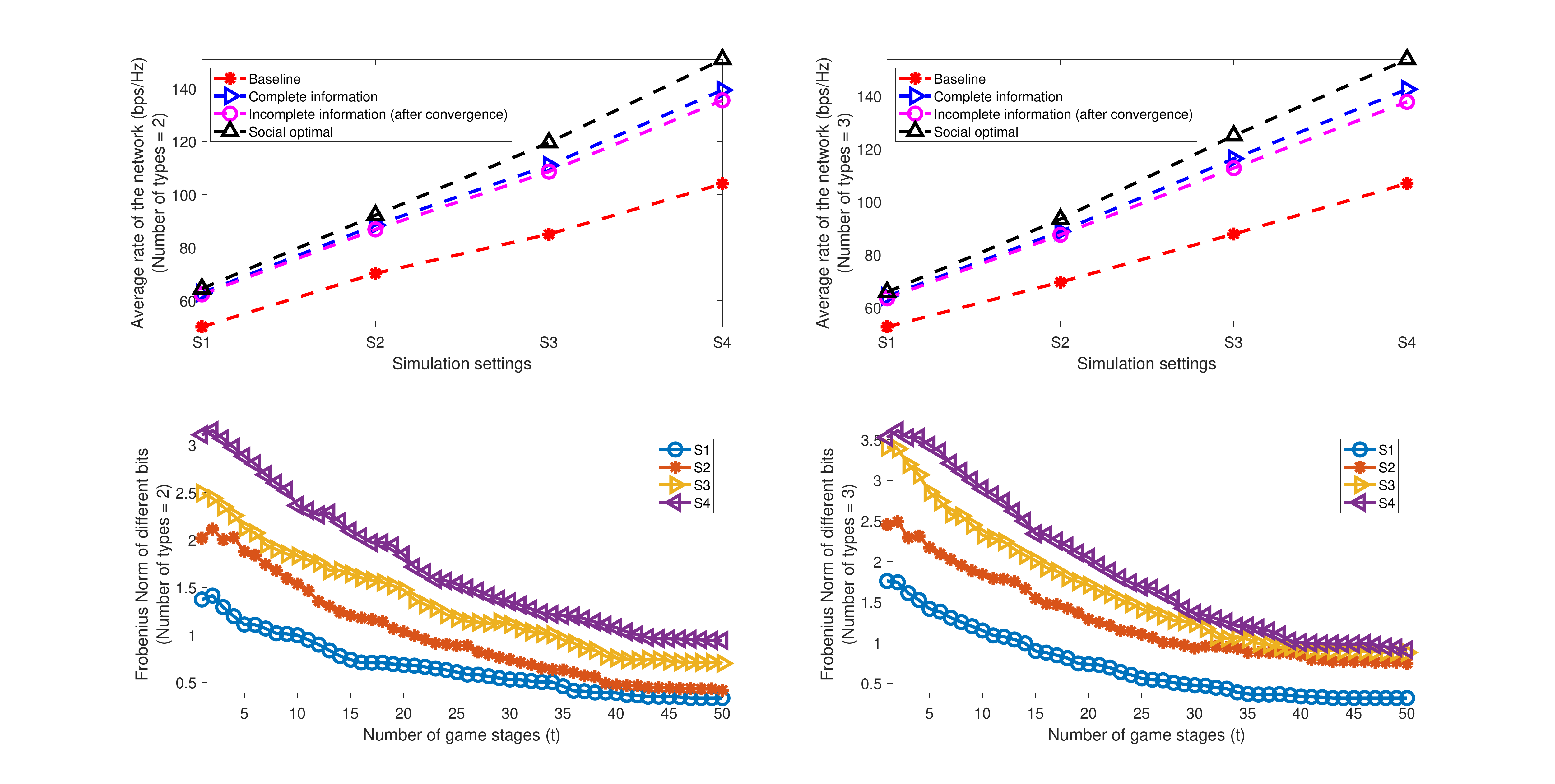}
\caption{Average transmission rate of the network for each simulation setting shown in Table~\ref{table:simulation_settings}. It shows the effect of increase in the overlap of the distributions. First case is for two types ($T_1: \mu=12, \sigma=6;\ T_2: \mu=18, \sigma=6$) and second case is for three types ($T_1: \mu=12, \sigma=6;\ T_2: \mu=18, \sigma=6;\ T_3: \mu=24, \sigma=6$).}
\label{fig:subplots_for_60_per_overlap}	
\end{figure*}

So far, we have selected the types in a way that the corresponding probability distributions overlap only very slightly or not at all. However, it is worth noting that in case of highly overlapping type distributions, the accuracy and the convergence rate would degrade. \textbf{Fig. \ref{fig:subplots_for_60_per_overlap}} illustrates the case where the types' distributions overlap approximately up to $60 \%$. 

As the final remark, we mention the following: The repeated coalition formation algorithms with incomplete and complete information yield greater average rates than the baseline configuration and lower than the social optimal case. In the social optimal configuration, the optimal coalition structure is the one that maximizes the overall rate of the network such that every drone provides a rate more than the baseline configuration. However, the socially optimal solution might not be a stable coalition structure and shall be enforced by an authority.  

\section{Conclusion} 
\label{conclusion}
For a drone-based wireless network, we have developed an approach for distributed cooperation among drones under uncertainty based on a Bayesian coalition formation process in order to maximize the overall transmission rate in the network.  Given only limited information about the {\em type} (i.e. amount of available power) in other drones, coalitions of drones are formed, the spectrum and energy resources are pooled, and then the channel allocations to users are shuffled to improve network performance. We have evaluated the method through theoretical and numerical analysis. Theoretical analysis has shown the convergence to the stable coalition structure. Also, simulation results have showed performance improvement over baseline method. Future works include generalization to the scenario with state-dependent types with unknown states as well as the scenario with high-dimensional types.   
\bibliographystyle{IEEEtran}
\bibliography{references}
\end{document}